\documentclass[a4paper,11pt]{article}
\pdfoutput=1
\usepackage{jcappub}
\usepackage{slashed}
\usepackage{mathtools}
\usepackage[T1]{fontenc}
\usepackage[normalem]{ulem}
\usepackage{float}
\usepackage{orcidlink}

\graphicspath{{plots/}}

\newcommand{\gs}{g_\star}

\newcommand{\Trh}{T_\text{rh}}
\newcommand{\ct}{c_\theta}
\newcommand{\st}{s_\theta}
\newcommand{\Gp}{\Gamma_\phi}

\title{Reheating the FCC:\\Probing Early Matter Domination with Long-Lived Particles}
\author[a]{Nicolás Bernal\ \orcidlink{0000-0003-1069-490X},}
\author[b,c]{Giovanna Cottin\ \orcidlink{0000-0002-5308-5808},}
\author[a]{\\Kuldeep Deka\,\orcidlink{0000-0002-0064-5488}}
\author[b,c]{and Manuel López\,\orcidlink{0009-0008-3248-3381}}

\affiliation[a]{New York University Abu Dhabi\\
PO Box 129188, Saadiyat Island, Abu Dhabi, United Arab Emirates}
\affiliation[b]{Instituto de Física, Pontificia Universidad Católica de Chile\\
Avenida Vicuña Mackenna 4860, Santiago, Chile}
\affiliation[c]{Millennium Institute for Subatomic Physics at the High Energy Frontier (SAPHIR)\\
Fernández Concha 700, Santiago, Chile}

\emailAdd{nicolas.bernal@nyu.edu}
\emailAdd{gfcottin@uc.cl}
\emailAdd{kuldeep.deka@nyu.edu}
\emailAdd{manuel.lopez.f@uc.cl}

\abstract{We study a GeV-scale Higgs-portal scalar $\phi$ that can dominate the early Universe and later decay into Standard Model states, ending an early matter-dominated era. Ordinary cosmic reheating is the minimal example of this scenario. The small Higgs mixing makes $\phi$ a long-lived particle (LLP), linking the reheating temperature $\Trh$ to displaced-decay signatures at colliders. Including finite-temperature suppression of the decay width, we map FCC-hh LLP sensitivity onto the $[m_\phi,\, \Trh]$ plane. We find that FCC-hh displaced searches could probe GeV-scale transition temperatures, up to the electroweak scale within the broken-phase Higgs-portal description.}
\begin{document}
\begin{flushright}
\end{flushright}
\maketitle

%%%%%%%%%%%%%%%%%%%%%%%%%%%%%%%%%%%%%
\section{Introduction}
%%%%%%%%%%%%%%%%%%%%%%%%%%%%%%%%%%%%%
The expansion history of the Universe before Big-Bang nucleosynthesis (BBN) is only weakly constrained. While the standard cosmological picture assumes radiation domination at early times, many well-motivated extensions of the Standard Model (SM) predict non-standard epochs prior to BBN~\cite{Allahverdi:2020bys, Batell:2024dsi}. A particularly simple possibility is an early matter-dominated era, driven by the energy density of a non-relativistic field or long-lived particle $\phi$. Such a period can arise from coherent scalar oscillations, moduli fields, hidden-sector states, or other massive particles whose lifetime is long compared with the Hubble time at early epochs. The minimal and most familiar example is cosmic reheating, where $\phi$ is identified with the inflaton or reheaton and its decay transfers the energy of the inflationary sector to the SM thermal bath~\cite{Dolgov:1989us, Traschen:1990sw, Kofman:1994rk, Kofman:1997yn, Barman:2025lvk}. More generally, however, the same dynamics describes any early period in which a long-lived matter component dominates the energy density and subsequently decays into SM radiation.

The temperature at which this matter-dominated era ends is a key parameter of the early thermal history. In this work, we denote it by $\Trh$, following the usual reheating terminology, but it should be understood more generally as the temperature of the SM bath at the transition from $\phi$-domination to SM radiation domination. In the special case where $\phi$ is the inflaton, $\Trh$ is the conventional post-inflationary reheating temperature. In more general scenarios, it is the decay or transition temperature of an early matter-dominated epoch. Low-temperature transitions are especially interesting because they point to feebly coupled particles with macroscopic lifetimes.

Long-lived particles (LLPs)~\cite{Alimena:2019zri} therefore provide a natural bridge between collider physics and the pre-BBN expansion history. A particle that dominates the early Universe must eventually decay into SM radiation. If its couplings to the SM are small, the same small decay width that delays the onset of radiation domination can also make the particle long-lived on collider scales. Conversely, a displaced decay observed at a collider would give direct information about the lifetime and couplings of a particle that may have controlled the early expansion rate. This connection motivates studying LLP signatures not only as evidence for new particles but also as probes of non-standard cosmological epochs.

We focus on a minimal Higgs-portal realization of this idea. The SM is extended by a real singlet scalar $\phi$ that mixes with the Higgs boson after electroweak symmetry breaking (EWSB). We remain agnostic about the mechanism that produced the primordial abundance of $\phi$ and simply assume that, at some early time, its energy density dominates over that of SM radiation. The scalar $\phi$ may be the inflaton/reheaton, but it may also represent a more general matter component such as a modulus-like field. Through its Higgs admixture, $\phi$ inherits SM-like interactions suppressed by the mixing angle $\theta$. The same mixing angle controls both its decay rate in the early Universe and its production rate at colliders.

For the GeV-scale masses studied here, decays of $\phi$ into pairs of Higgs bosons are kinematically closed. The scalar instead decays through its Higgs mixing into the heaviest kinematically accessible SM states, dominantly fermions in the parameter space of interest. In the early Universe these decays occur in a thermal plasma rather than in vacuum. For a non-relativistic scalar decaying into thermalized fermions, Pauli blocking and inverse decays modify the decay width according to~\cite{Adshead:2019uwj}
\[
    \Gp(T) = \Gp^{(0)}\, \tanh\!\left(\frac{m_\phi}{4\, T}\right),
\]
where $\Gp^{(0)}$ corresponds to its total decay width in vacuum. At temperatures $T \gtrsim m_\phi/4$, this in-medium rate is suppressed relative to the vacuum width. This thermal effect changes the relationship between the lifetime of $\phi$ and the transition temperature $\Trh$, and is therefore important when mapping collider-accessible LLP lifetimes onto early-Universe histories.

Collider searches for displaced decays provide a direct probe of this setup. However, at the LHC, the production rate of $\phi$ is suppressed by the same small mixing angle that makes it long-lived. In addition, existing displaced-vertex searches are not optimized for light GeV-scale scalars with relatively soft visible decay products. This motivates the study of future colliders with higher production rates and improved LLP coverage. The Future Circular Collider in hadron-hadron mode (FCC-hh)~\cite{FCC:2018vvp, Benedikt:2022kan, FCC:2025uan}, with a proposed center-of-mass energy $\sqrt{s} = 100$~TeV and integrated luminosity of the order of 20~ab$^{-1}$, offers a particularly promising environment. We consider the proposed FCC-hh central and forward trackers, as well as the dedicated forward detector concept FOREHUNT~\cite{Bhattacherjee:2023plj}.

The goal of this work is to connect displaced-decay sensitivity at future colliders with the temperature at which an early matter-dominated era ends. We compute $\Trh$ as a function of the scalar mass and the Higgs mixing angle, including the finite-temperature suppression of the decay width. We then simulate the production and displaced decays of $\phi$ at colliders and estimate the reach of several FCC-hh detector concepts. The result is a direct map between LLP searches in the $[m_\phi,\, \sin\theta]$ plane and cosmological targets in the $[m_\phi,\, \Trh]$ plane. In minimalist interpretation, this map probes low-temperature cosmic reheating. In the more general interpretation emphasized here, it probes the end of a pre-BBN early matter-dominated epoch.

The paper is organized as follows. In Section~\ref{sec:framework} we introduce the Higgs-portal scalar model and define the mass eigenstates and couplings. In Section~\ref{sec:cosmo} we compute the temperature at which the $\phi$-dominated era ends, emphasizing the role of finite-temperature effects in the decay width. In Section~\ref{sec:colliders} we discuss the LLP phenomenology at colliders, including the limited LHC reach and the projected sensitivity of FCC-hh detector concepts. We then overlay the collider reach with the contours of $\Trh$, showing which early matter-dominated histories could be tested. We conclude in Section~\ref{sec:concl}. Details of the finite-temperature decay calculation and of the relevant LLP kinematics are collected in the appendices.

%%%%%%%%%%%%%%%%%%%%%%%%%%%%%%%%%%%%%
\section{The set-up} \label{sec:framework}
%%%%%%%%%%%%%%%%%%%%%%%%%%%%%%%%%%%%%
We consider a minimal model where the SM is enlarged with a real singlet scalar $\tilde\phi$ that only couples through a trilinear interaction with the Higgs doublet $\Phi$~\cite{Cado:2025orb}. Before EWSB, the scalar potential is given by
\begin{equation} \label{eq:potential_unbroken}
    V = \mu_\Phi^2\, |\Phi|^2 + \lambda_\Phi\, |\Phi|^4 + \frac12\, m_{\tilde \phi}^2\, \tilde \phi^2 + \mu\, \tilde\phi\, |\Phi|^2,
\end{equation}
with $\mu$ a mass-dimension coupling. For phenomenological purposes, we neglect the possible quadratic term $\tilde \phi^2\, |\Phi|^2$ since we will focus on decays and not annihilations; in this way, this is a truncated effective model. In addition, trilinear and quartic interactions of $\tilde\phi$ are also ignored, as they do not play a major role in the upcoming analysis. Finally, we assume that the omitted UV sector stabilizes the potential.
 
After the EWSB, the Higgs takes a vacuum expectation value $v \simeq 246$~GeV, and, in the unitary gauge where $\Phi \to \left(0,\, (v + \tilde h)/\sqrt{2}\right)^T$, the potential becomes
\begin{equation}
    V = \frac{m_{\tilde h}^2}{2}\, \tilde h^2 + \lambda_\Phi\, v\, \tilde h^3 + \frac{\lambda_\Phi}{4}\, \tilde h^4 + \frac{m_{\tilde \phi}^2}{2}\, \tilde \phi^2 + \mu\, v\, \tilde \phi\, \tilde h + \frac{\mu}{2}\, \tilde \phi\, \tilde h^2,
\end{equation}
where we have used $v^2 = - \mu_\Phi^2/\lambda_\Phi$, $m_{\tilde h}^2 \equiv 2\, \lambda_\Phi\, v^2$, and have absorbed the linear term in $\tilde\phi$ with a field redefinition. To diagonalize the mass matrix for $\tilde h$ and $\tilde \phi$, one should redefine the fields as
\begin{align}
    \tilde h    &= + h\, \cos\theta + \phi\, \sin\theta\,, \label{eq:mixing1}\\
    \tilde \phi &= - h\, \sin\theta + \phi\, \cos\theta\,, \label{eq:mixing2}
\end{align}
in terms of the field $h$ and $\phi$ rotated by an angle $\theta$ given by
\begin{equation}
   \tan 2\theta = \frac{2\,\mu\, v}{m_{\tilde\phi}^2 - m_{\tilde h}^2}\,.
\end{equation}
Therefore, the scalar potential becomes
\begin{align}
    V &= \frac{m_h^2}{2} h^2 + \left(\lambda_\Phi\, \ct^3\, v - \frac{\ct^2\, \st}{2} \mu\right) h^3 + \frac{\lambda_\Phi\, \ct^4}{4} h^4 + \frac{m_\phi^2}{2} \phi^2 + \left(\lambda_\Phi\, \st^3\, v + \frac{\ct\, \st^2}{2} \mu\right) \phi^3 + \frac{\lambda_\Phi\, \st^4}{4} \phi^4 \nonumber\\
    &\quad + \frac{\st}{4} \left(+\mu + 3\, c_{2\theta}\, \mu + 6\, s_{2\theta}\, \lambda_\Phi\, v\right) h\, \phi^2 + \frac{\ct}{4} \left(-\mu + 3\, c_{2\theta}\, \mu + 6\, s_{2\theta}\, \lambda_\Phi\, v\right) \phi\, h^2\nonumber\\
    &\quad + \lambda_\Phi\, \ct\, \st^3\, h\, \phi^3 + \lambda_\Phi\, \ct^3\, \st\, \phi\, h^3 + \frac{3\, \ct^2\, \st^2\, \lambda_\Phi}{2}\, \phi^2\, h^2,
\end{align}
with $c_x \equiv \cos(x)$, $s_x \equiv \sin(x)$, and
\begin{align}
    m_h^2 &\equiv \frac12 \left[m_{\tilde h}^2 + m_{\tilde\phi}^2 + \sqrt{\left(m_{\tilde h}^2 - m_{\tilde\phi}^2\right)^2 + 4\, v^2\, \mu^2}\right],\\
    m_\phi^2 &\equiv \frac12 \left[m_{\tilde h}^2 + m_{\tilde\phi}^2 - \sqrt{\left(m_{\tilde h}^2 - m_{\tilde\phi}^2\right)^2 + 4\, v^2\, \mu^2}\right].
\end{align}
The heavier state $h$ is identified with the SM-like Higgs boson with mass $m_h \simeq 125$~GeV, while the lighter state $\phi$ is identified with the inflaton/reheaton or, in general, the particle that drove the evolution of the Universe.

Mixing with $\phi$ modifies the couplings of the Higgs boson $h$ with the SM particles. From Eqs.~\eqref{eq:mixing1} and~\eqref{eq:mixing2} it can be deduced that its couplings with SM particles will be suppressed by a factor $\cos\theta$. In addition, the Higgs-like particle could also decay into a pair of $\phi$, with a coupling suppressed by $\sin\theta$. Therefore, small mixing angles are required to keep Higgs physics in agreement with high-precision measurements. We have verified with \texttt{HiggsTools}~\cite{Bahl:2022igd} that for $\sin\theta \ll 0.1$ the model is compatible with current LHC data from Higgs physics.

%%%%%%%%%%%%%%%%%%%%%%%%%%%%%%%%%%%%%
\section{Higgs-portal decay and the transition to radiation domination} \label{sec:cosmo}
\begin{figure}[t!]
    \def\sepf{0.49}
    \centering
    \includegraphics[width=0.6\columnwidth]{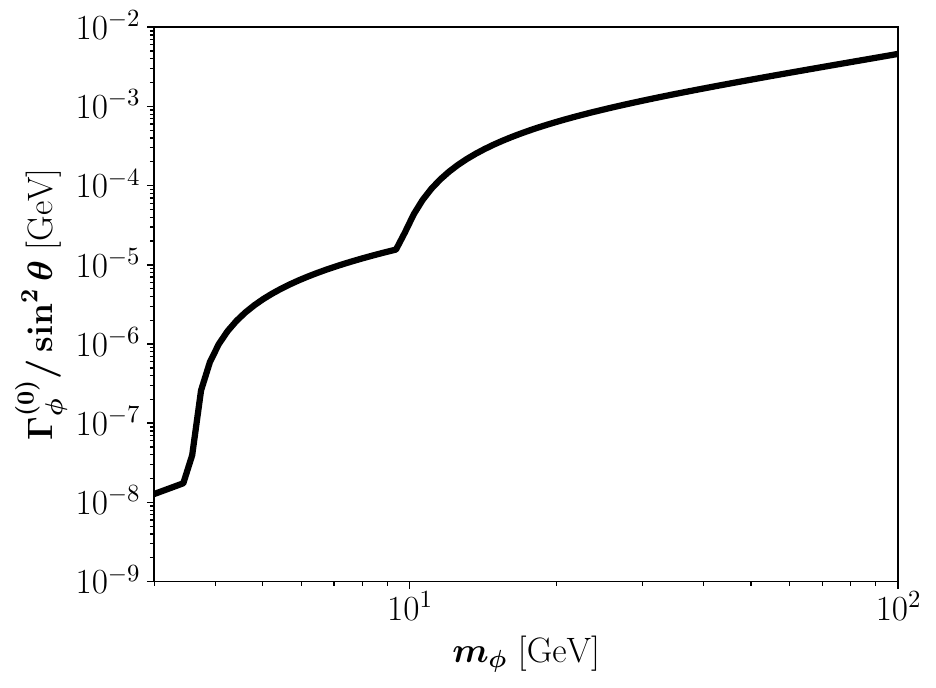}
    \caption{Total decay width $\Gp^{(0)}$ of $\phi$ divided by $\sin^2\theta$, as a function of its mass $m_\phi$.}
    \label{fig:dec}
\end{figure}

%%%%%%%%%%%%%%%%%%%%%%%%%%%%%%%%%%%%%
The {\it in-vacuum} decay width $\Gp^{(0)}$ of $\phi$ is
\begin{equation} \label{eq:phidecay}
    \Gp^{(0)} = \sin^2\theta\, \Gamma_h^\text{SM}(m_\phi) + \Gamma_{\phi \to h h} + \Gamma_{\phi \to h h h}
\end{equation}
and contains three contributions. The first corresponds to the decays into fermions and gauge bosons, and is the same as the one for the SM Higgs boson (with mass $m_\phi$) suppressed by the square of the sine of the mixing angle $\theta$. The second and third terms correspond to the decay into two and three Higgs bosons, respectively. In our setup $m_\phi$ lies below the electroweak scale, so these decays are always kinematically closed, and $\phi$ decays dominantly into SM fermions. Figure~\ref{fig:dec} shows the total decay width of $\phi$ divided by $\sin^2\theta$, as a function of $m_\phi$; it was produced with \texttt{MadGraph5}~\cite{Alwall:2011uj, Alwall:2014hca}. The thresholds around $m_\phi \simeq 3.5$~GeV and $m_\phi \simeq 10$~GeV correspond to the opening of the decay channels to pairs of $\tau$ leptons and $b$ quarks, respectively~\cite{Djouadi:2005gi}. Below the $b$-quark threshold, the Higgs-like scalar width is subject to hadronic uncertainties. We used the MadGraph partonic width as a benchmark estimate and left a treatment with dedicated light-scalar hadronic widths for future work.
%%%%%%%%%%%%%%%%%%%%%%%%%%%%%%%%%%%%%%%%%%%%%%%%%%%
%%%%%%%%%%%%%%%%%%%%%%%%%%%%%%%%%%%%%%%%%% 

\begin{figure}[t!]
    \def\sepf{0.496}
    \centering
    \includegraphics[width=\sepf\columnwidth]{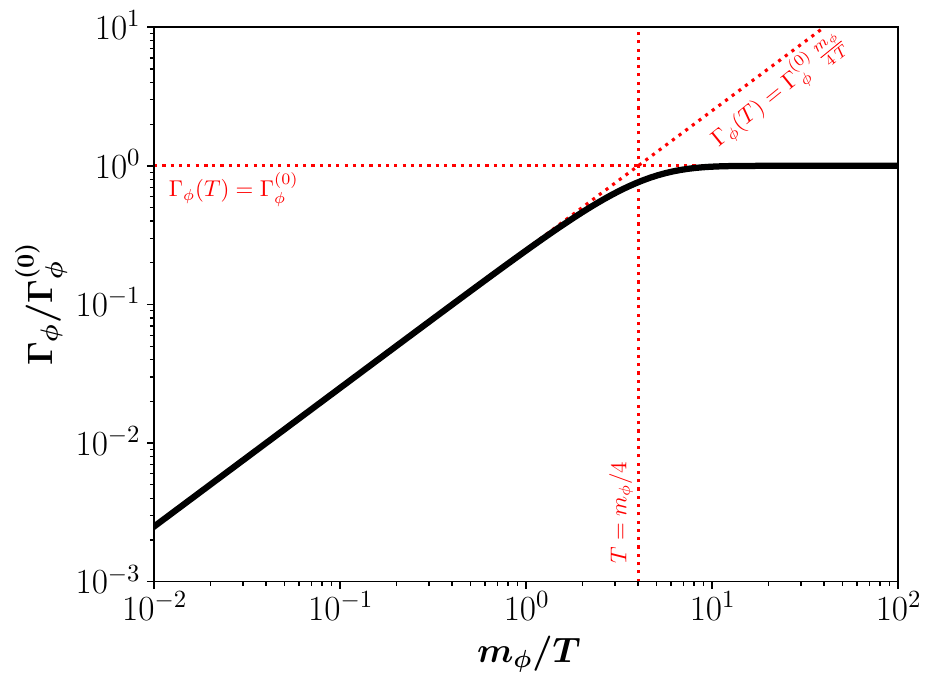}
    \includegraphics[width=\sepf\columnwidth]{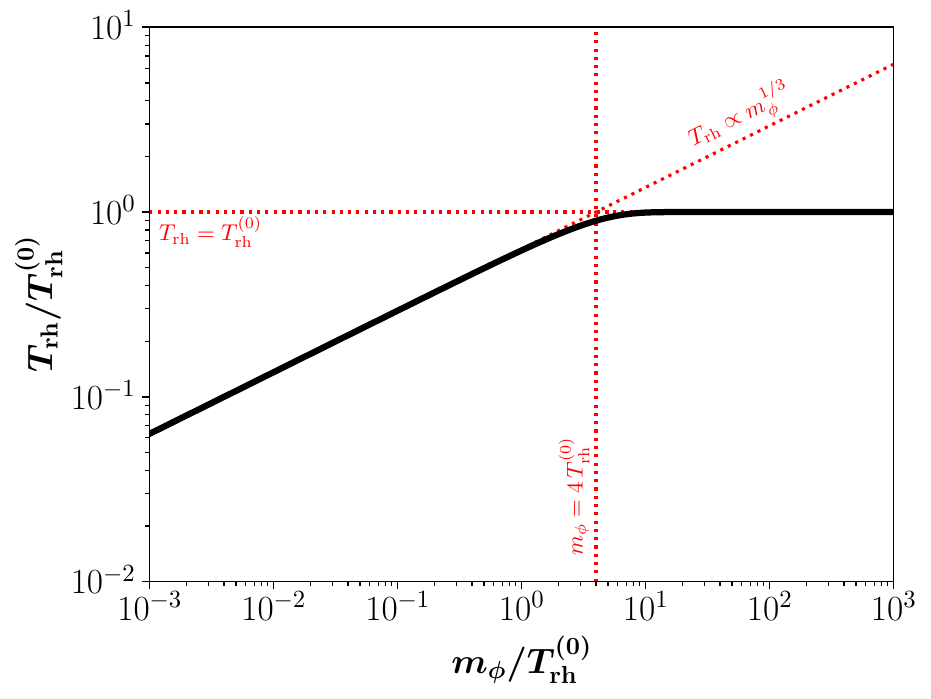}
    \caption{Variation of the temperature-dependent decay width $\Gp(T)$ with respect to its zero-temperature value $\Gp^{(0)}$, as a function of the $m_\phi/T$ (left panel). The right panel shows the corresponding impact on the reheating temperature $\Trh$.}
    \label{fig:Geff}
\end{figure}

However, in the early Universe $\phi$ does not decay in vacuum, but in the presence of a SM thermal plasma, which can have a strong impact on its cosmic evolution. The quantum statistics of non-relativistic bosons $\phi$ that decay into SM fermions in thermal equilibrium with the SM thermal plasma at a temperature $T$ modifies the decay width $\Gp$ of $\phi$ with respect to its value in vacuum $\Gp^{(0)}$, and renders it temperature dependent~\cite{Adshead:2019uwj}
\begin{equation}  \label{eq:Gexact}
    \Gp(T) = \Gp^{(0)}\, \tanh\left(\frac{m_\phi}{4\, T}\right),
\end{equation}
which can be approximated as 
\begin{equation} \label{eq:Gapprox}
    \Gp(T) \simeq \Gp^{(0)} \times
    \begin{dcases}
        1 &\text{ for }T \leq \frac{m_\phi}{4}\,,\\
        \frac{m_\phi}{4\, T} &\text{ for }T \geq \frac{m_\phi}{4}\,,
    \end{dcases}
\end{equation}
see Appendix~\ref{sec:decay} for a detailed derivation. For temperatures below $m_\phi/4$, the thermal bath cannot produce a sizable backreaction, and the in-vacuum approach works well. However, for $T > m_\phi/4$, its decay width is suppressed, as shown in the left panel of Fig.~\ref{fig:Geff}. Strictly speaking, Eq.~\eqref{eq:Gexact} applies channel by channel to decays into fermions in thermal equilibrium; this is the case in the present scenario, since the total width of $\phi$ is dominated by SM fermions in the final states.
%%%%%%%%%%%%%%%%%%%%%%%%%%%%%%%%%%%%%%%%%%%%%%%%%%%
%%%%%%%%%%%%%%%%%%%%%%%%%%%%%%%%%%%%%%%%%% 

The energy density stored in $\phi$ could have dominated the total energy density of the early Universe. This is our working assumption. This scenario naturally arises if $\phi$ is identified with the inflaton (or the reheaton) during the cosmic reheating era~\cite{Dolgov:1989us, Traschen:1990sw, Kofman:1994rk, Kofman:1997yn, Allahverdi:2020bys, Batell:2024dsi, Barman:2025lvk}. However, that could also occur in more general scenarios where a long-lived heavy particle $\phi$ generates an early matter-dominated epoch, like in the case of moduli fields from string compactification~\cite{Ellis:1986zt, Banks:1993en, Kane:2015jia, Cicoli:2023opf}. To be consistent with the BBN observations, $\phi$ has to decay, leading to an era dominated by SM radiation.

The reheating temperature $\Trh$ corresponds to the temperature of the SM plasma at the moment of the equality of the energy densities of the SM and $\phi$. A precise  determination of $\Trh$ requires solving the coupled Boltzmann equations for the energy densities. In this work, we use the standard analytic estimate $H(\Trh) = \Gp(\Trh)$, which captures the parametric dependence on $m_\phi$ and $\sin\theta$, where the Hubble expansion rate $H$ for a universe dominated by SM radiation is given by
\begin{equation} \label{eq:H}
    H(T) = \frac{\pi}{3} \sqrt{\frac{\gs(T)}{10}}\, \frac{T^2}{M_P}\,,
\end{equation}
 where $\gs$ corresponds to the number of relativistic degrees of freedom that contribute to the SM energy density~\cite{Drees:2015exa}, and $M_P \simeq 2.4 \times 10^{18}$~GeV is the reduced Planck mass. Therefore, using the approximations in Eq.~\eqref{eq:Gapprox},
\begin{equation}
    \Trh \simeq
    \begin{dcases}
        \left(\frac{3}{\pi}\right)^{1/2} \left(\frac{10}{\gs}\right)^{1/4} \left(M_P\, \Gp^{(0)}\right)^{1/2} &\text{ for } m_\phi \geq 4\, \Trh\,,\\
        \left(\frac{3}{4\pi}\right)^{1/3} \left(\frac{10}{\gs}\right)^{1/6} \left(M_P\, \Gp^{(0)}\, m_\phi\right)^{1/3} &\text{ for } m_\phi \leq 4\, \Trh\,,
    \end{dcases}
\end{equation}
as presented in the right panel of Fig.~\ref{fig:Geff}. Upon comparing the two panels, it can be seen that the suppression of the decay width is reflected in the decrease in the reheating temperature compared to the case where the decay occurs in vacuum.

\begin{figure}[t!]
    \def\sepf{0.49}
    \centering
    \includegraphics[width=0.6\columnwidth]{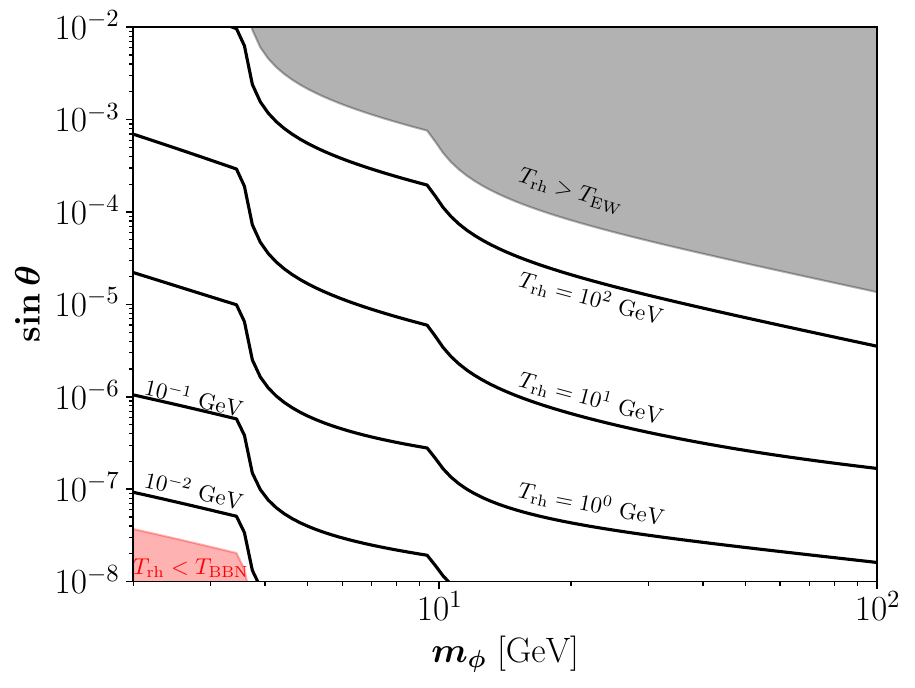}
    \caption{Contours for the reheating temperature $\Trh$ as a function of the mass $m_\phi$ and the mixing angle $\theta$. We exclude the regions for which $\Trh < T_\text{BBN}$ and $\Trh > T_{\rm EW}$.}
    \label{fig:Trh}
\end{figure}

In our particular scenario, the reheating temperature can be calculated for given values of the mass $m_\phi$ and the mixing angle $\theta$ using Eqs.~\eqref{eq:phidecay}, \eqref{eq:Gexact} and~\eqref{eq:H}. Alternatively, the mixing angle can also be calculated as a function of $m_\phi$ and $\Trh$:
\begin{equation} \label{eq:sintheta_Trh_thermal}
    \sin^2\theta = \frac{\pi}{3} \sqrt{\frac{\gs(\Trh)}{10}}\, \frac{\Trh^2}{M_P\, \Gamma_h^{\rm SM}(m_\phi)}\, \coth\left(\frac{m_\phi}{4\,\Trh}\right).
\end{equation}
Figure~\ref{fig:Trh} shows the contours for $\Trh$ in the plane $[m_\phi,\, \sin\theta]$. For small mixing values, $\Trh$ is very suppressed and can be even smaller than $T_\text{BBN} \simeq 4$~MeV~\cite{Sarkar:1995dd, Kawasaki:2000en, Hannestad:2004px, DeBernardis:2008zz, deSalas:2015glj}. This region (red area) is excluded to avoid spoiling the success of BBN. In contrast, high values of the mixing angle could produce a decay that occurs mostly before the EWSB (gray area). Even if viable, in this analysis, we are not interested in this possibility as will be seen in Fig.~\ref{fig:all}. The kinks around $m_\phi \simeq 3.5$~GeV and $m_\phi \simeq 10$~GeV correspond to the kinematical thresholds discussed in Fig.~\ref{fig:dec}.
%%%%%%%%%%%%%%%%%%%%%%%%%%%%%%%%%%%%%%%%%%%%%%%%%%%
%%%%%%%%%%%%%%%%%%%%%%%%%%%%%%%%%%%%%%%%%% 

%%%%%%%%%%%%%%%%%%%%%%%%%%%%%%%%%%%%%
\section{LLP phenomenology at colliders} \label{sec:colliders}
%%%%%%%%%%%%%%%%%%%%%%%%%%%%%%%%%%%%%
To study LLP phenomenology at colliders, we implemented the model in a UFO file with \texttt{Feynrules}~\cite{Christensen:2008py, Alloul:2013bka}. With this model file, we then simulate  production in $p p$ collisions with \texttt{MadGraph5}~\cite{Alwall:2011uj, Alwall:2014hca}. The particle $\phi$ and the Higgs boson have the same predominant production mechanisms as the SM Higgs boson; however, their couplings are suppressed by a factor of $\sin\theta$ and $\cos\theta$, respectively. Therefore, for their production, the channel that yields the highest cross-section is gluon-gluon fusion (ggF). We implemented this using an effective vertex via the following standard term~\cite{Anastasiou:2016cez}
\begin{align} \label{eq:EFT_vertex}
        \mathcal{L}_{ggF} &= \frac{c^{(1)}_h}{\Lambda}\, \cos\theta\, h\, G^{\mu\nu} G_{\mu\nu} + \frac{c^{(1)}_\phi}{\Lambda}\, \sin\theta\, \phi\, G^{\mu\nu} G_{\mu\nu}\,, \\
        \frac{c^{(1)}_X}{\Lambda} &\equiv \frac{\alpha_s}{12\pi\, v} \left[1+\frac{7}{120} \left(\frac{m_X}{m_t}\right)^2 + \frac{1}{168} \left(\frac{m_X}{m_t}\right)^4 + \frac{13}{16800} \left(\frac{m_X}{m_t}\right)^6\right],
\end{align}
where the Wilson coefficients $c^{(1)}_X$ were found by analytically calculating the one-loop diagrams. In the heavy top-quark limit ($m_t\to\infty$), this EFT coefficient yields similar results~\cite{Harlander:2000mg, Grigo:2014jma}. Additionally, we validated our implementation by comparing the ggF, the vector boson fusion (VBF) and total production cross-sections in Ref.~\cite{LHCHiggsCrossSectionWorkingGroup:2011wcg} with those obtained in \texttt{MadGraph5} with our model setting $\theta = 0$, in order to reproduce the SM couplings. 

%%%%%%%%%%%%%%%%%%%%%%%%%%%%%%%%%%%%%%%%%%%%%%%%%%%
\begin{figure}[t!]
    \def\sepf{0.49}
    \centering
    \includegraphics[width=\sepf\columnwidth]{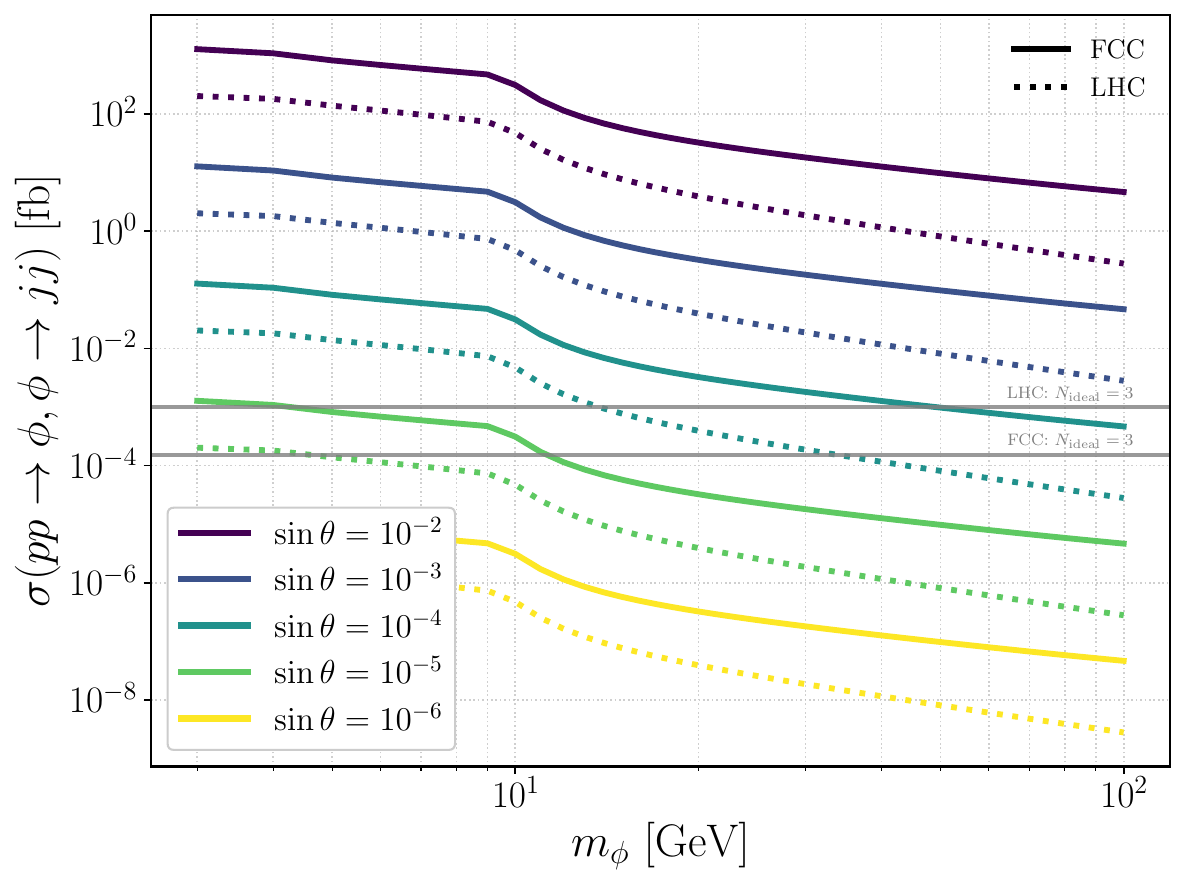}
    \includegraphics[width=\sepf\columnwidth]{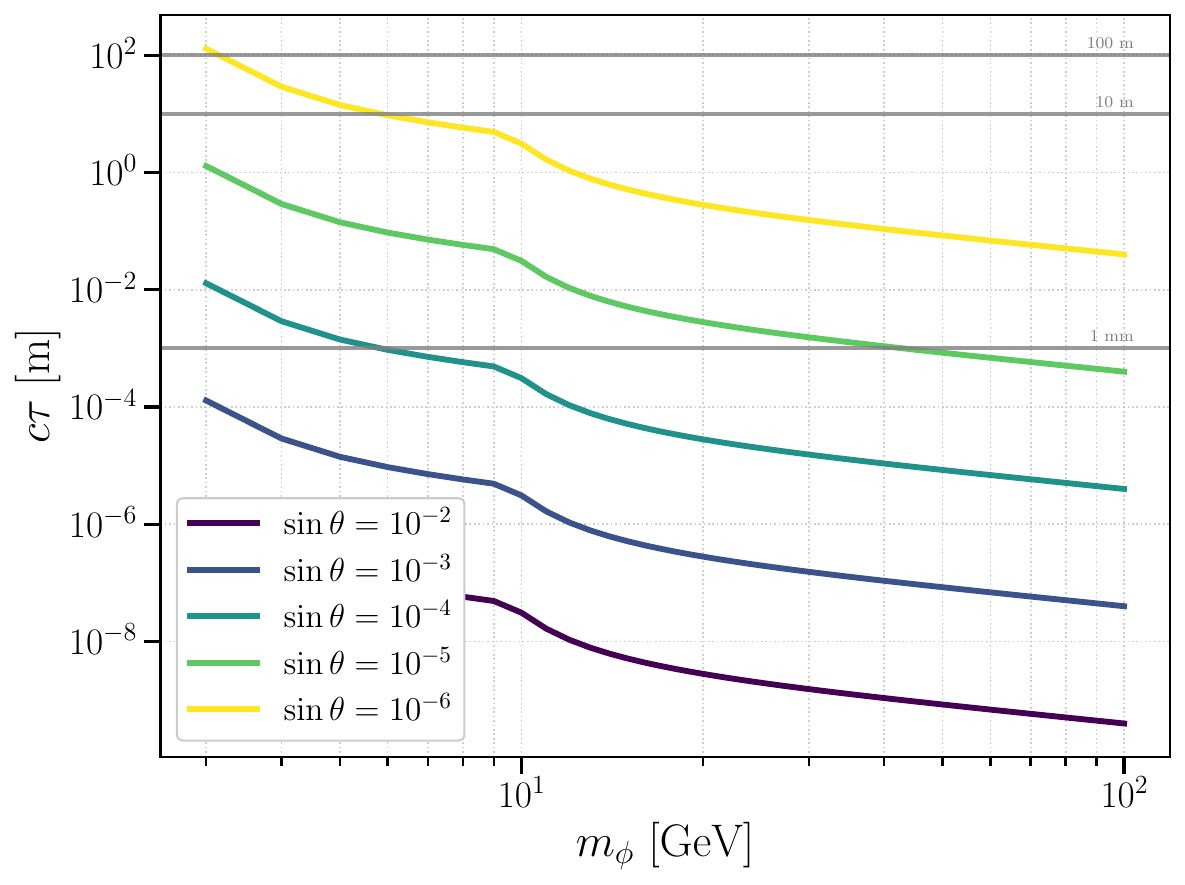}
    \caption{Production cross-section at $\sqrt{s}=14$~TeV (LHC) and  $\sqrt{s}=100$~TeV (FCC) as a function of inflaton mass (left). A requirement of $N_{\text{ideal}} = \sigma(pp \rightarrow \phi) \times \text{BR}(\phi \to jj) \times \mathcal{L}$ of 3 events is shown in horizontal lines at $\mathcal{L} = 3$~ab$^{-1}$ for the LHC and $\mathcal{L} = 20$~ab$^{-1}$ at FCC. Curves of proper decay distance $c\tau=c\hbar/\Gamma_{\phi}$ (right) as a function of inflaton mass, for several fixed values of $\sin{\theta}$. }
    \label{fig:prod-dec}
\end{figure} 
%%%%%%%%%%%%%%%%%%%%%%%%%%%%%%%%%%%%%%%%%% 
We generate events at  $\sqrt{s} = 14$~TeV for the LHC and $\sqrt{s} = 100$~TeV for the FCC in \texttt{MadGraph5} with our model implementation. The left panel of Figure~\ref{fig:prod-dec} shows the production cross-sections and the decay of $\phi$ to jets, as a function of mass, for several fixed values of $\sin{\theta}$. The horizontal lines in this figure represent an ideal (i.e. no detector acceptance or efficiencies) number of 3 expected events, $N_{\text{ideal}}=\sigma(p p \to \phi)\times \text{BR}(\phi \to jj) \times \mathcal{L}$, at both the LHC (with $\mathcal{L}=3000$ fb$^{-1}$) and FCC-hh (with $\mathcal{L}=20$ ab$^{-1}$). The right frame in Fig.~\ref{fig:prod-dec} shows the proper decay distance, $c\tau$, of $\phi$ as a function of its mass, for different fixed choices of the mixing. These plots help us to understand the sensitivity region at the LHC and FCC-hh in the model parameter space to be targeted. With cross-section alone, it is not possible to access values of $\sin{\theta}$ lower than $10^{-5}$, as these will fall below $N_{\text{ideal}}$. Lower values at $10^{-6}$ and below maximize the value of $c\tau$ between 10 and 100 meters, but these would be undetectable due to low cross-sections. We checked the potential sensitivity with proposed dedicated far and transverse detectors at the LHC such as MATHUSLA or ANUBIS, with the \texttt{Displaced Decay Counter (DDC)} program~\cite{Domingo:2023dew}, which computes decay probabilities at several proposed detectors. We get no visible long-lived signal yield at the proposed LHC detectors. This is mainly due to the suppression factor of $\sin{\theta}$. Smaller values of $\sin\theta$ enhance $c\tau$ but in turn suppress the production cross-section, and therefore very low rates are expected at the LHC. In addition, current displaced vertex (DV) searches at the main detectors such as ATLAS~\cite{ATLAS:2023oti} rely on cuts on the invariant mass of the displaced vertex to be larger than 10~GeV, and therefore there is limited sensitivity in our region of interest, for $\mathcal{O}(10)$~GeV inflaton masses. Further cuts in the transverse and longitudinal displacement from the interaction point (IP) of minimum $4$~mm and maximum of 300 mm are required to be within the acceptance of the ATLAS inner tracker, where displaced vertices can be reconstructed. Strong kinematic cuts on jets~\cite{ATLAS:2023oti} also reduce the sensitivity to our $\mathcal{O}(10)$~GeV inflaton;\footnote{A recent recast of the ATLAS DV + jets search~\cite{ATLAS:2023oti} was done in Ref.~\cite{Wang:2024ieo} for a Higgs-portal model with a light scalar. But in that case, the displaced signature came from exotic Higgs decays, and production was from a SM Higgs. In our case, production is suppressed by mixing.} see Appendix~\ref{sec:appendix} for details on the impact of these cuts in our model. All in all, we conclude that the LHC is not optimal to access the LLP phenomenology of the model.\footnote{See our previous work in Ref.~\cite{Bernal:2025qkj} for a similar conclusion in a dark matter Higgs-portal model.} We note that since the production cross section of Higgs-like particles is tightly related to the Yukawa couplings, we would expect a significant suppression of the number of expected events at a leptonic collider. Therefore, as the LHC has limited sensitivity to our model at the expected number of events level, we would expect the FCC-ee to be even less useful for this task. A dedicated study of FCC-ee is beyond the scope of this work. Here we focus on FCC-hh because ggF production at 100~TeV provides the largest event rates for the Higgs-mixed scalar in the parameter region considered.

The Future Circular Collider (FCC) in hadronic mode (FCC-hh)~\cite{FCC:2018vvp, FCC:2025uan} is proposed to run at a center-of-mass energy of $\sqrt{s} = 100$~TeV and to have an integrated luminosity of $\mathcal{L} = 20$~ab$^{-1}$, which, in contrast to the LHC parameters ($\sqrt{s} = 14$~TeV and $\mathcal{L} = 3$~ab$^{-1}$), effectively yields roughly two orders of magnitude higher in sensitivity, making it a significantly more promising collider for our model region (see Fig.~\ref{fig:prod-dec}). We analyze our Monte Carlo events and compute $\sigma(p p \to \phi)$ and $\text{BR}(\phi \to f \bar{f})$, with $f$ being any SM charged fermion. The branching ratio remains close to one across $m_{\phi}$. Our $\texttt{MadGraph}$ events are also further analyzed with the \texttt{Displaced Decay Counter (DDC)} program~\cite{Domingo:2023dew}, which includes detector proposals for FCC-hh~\cite{C:2026bqd}.  Based on our model kinematics, we focus on three detectors:\footnote{We also computed estimations at DELIGHT~\cite{Bhattacherjee:2023plj, Bhattacherjee:2025gwo}, proposed as a dedicated transverse detector, but these are not shown as we note that the sensitivity for $c \tau \lesssim 10$~m is very limited, and not comparable to FOREHUNT, which performs much better.}
\begin{enumerate}
    \item FCC Central tracker: Corresponds to the main tracker of the proposed FCC-hh reference detector. It is an ATLAS-like cylindrical detector with a length $L_D = 5$~m starting from the IP, and an outer radius $R_O = 1.7$~m~\cite{FCC:2018vvp}.
    \item FCC Forward tracker: Similar to the Central tracker, but displaced $10$~m in the direction of the beam pipe from the IP, and slightly longer and narrower with a detector length of $L_D = 6$~m and an outer radius of $R_O = 1.6$~m~\cite{FCC:2018vvp}.
    \item FOREHUNT: Dedicated long-lived forward detector, similar to a perspective projection of the forward tracker along the $z$-axis. Its dimensions are $L_D = 50$~m and $R_O=5$~m. The distance from the IP is 50~m~\cite{Bhattacherjee:2023plj}.
\end{enumerate}

The probability of decay or acceptance $\epsilon$ of $\phi$ to decay inside a given detector is determined by reading the kinematic information of the events produced. This calculation generally follows
\begin{equation}
    \epsilon = e^{-\frac{L_1}{\beta\, \gamma\,  c\tau}} \left(1 - e^{-\frac{L_2 - L_1}{\beta\, \gamma\, c\tau}}\right),
\end{equation}
where
\begin{equation}
    \beta\, \gamma = \frac{|\vec{p}|}{E} \times \frac{E}{m_\phi} = \frac{|\vec{p}|}{m_\phi}\,,
\end{equation}
where $L_1$ and $L_2$ represent the path lengths along the LLP trajectory required to enter and exit the detector, respectively. These quantities depend on the polar angle $\Theta = \arctan(p_T/|p_z|)$ and on the detector scales $L_D$ and $R_O$. Dedicated formulas for a given detector can be found in the \texttt{DDC} github repository~\cite{DDcgithub}.

%%%%%%%%%%%%%%%%%%%%%%%%%%%%%%%%%%%%%%%%%%%%%%%%%%%%%%%%%%
\begin{figure}[t!]
    \centering
    \includegraphics[width=0.7\textwidth]{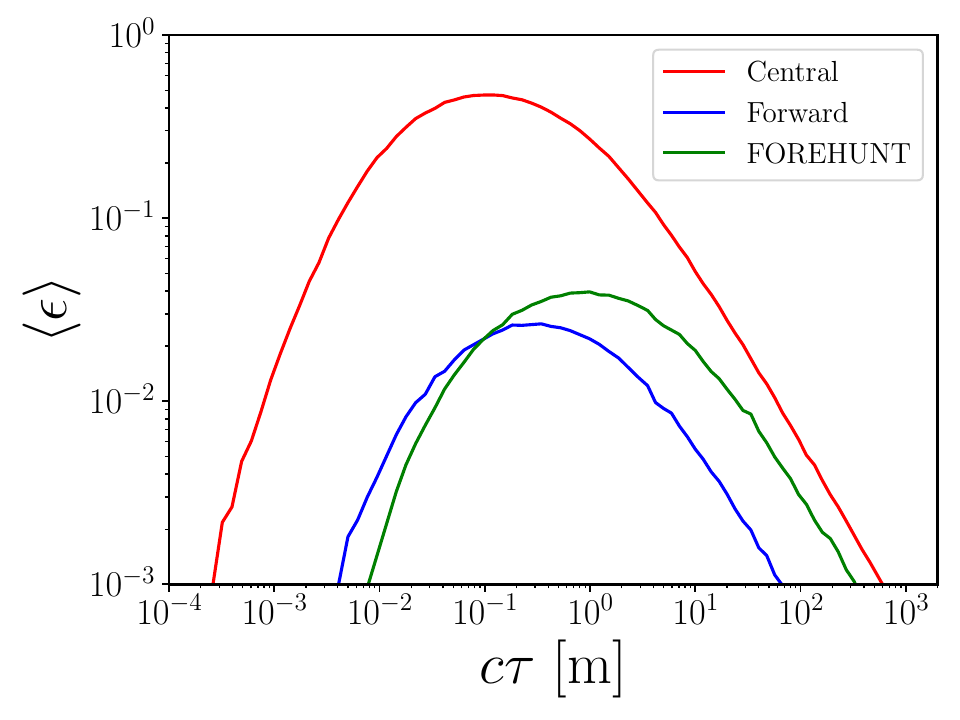}
    \caption{Average detector acceptance $\langle\epsilon\rangle$ as a function of the proper decay length $c \tau$, assuming $m_\phi = 5$~GeV, for the Central tracker, the Forward tracker and FOREHUNT.
    }
    \label{fig:detector_accept1}
\end{figure}
%%%%%%%%%%%%%%%%%%%%%%%%%%%%%%%%%%%%%%%%%%%%%%%%%%%%%%%%%%
Figure~\ref{fig:detector_accept1} shows the average acceptance $\langle\epsilon\rangle$ over all Monte Carlo simulated events for a representative benchmark with $m_\phi = 5$~GeV. The scanning of the displacement $c \tau$ comes from varying the mixing angle in the range $10^{-7} \lesssim \sin\theta \lesssim 10^{-2}$. For this range of parameters, the maximum average acceptance is $\sim 0.03$, $\sim 0.05$ and $\sim 0.5$, for the Forward tracker, FOREHUNT, and the Central tracker, respectively, in the range between 10~mm $\lesssim c \tau \lesssim 1$~m. The maximal acceptance is achieved in the Central tracker due to its large fiducial volume and undisplaced position relative to the IP, producing a smaller acceptance suppression compared to the displaced detectors. As expected, a higher mass means less kinetic energy is available, therefore a shorter decay distance in the lab frame, which results in a reduced acceptance.

%%%%%%%%%%%%%%%%%%%%%%%%%%%%%%%%%%%%%%%%%%%%%%%%%%%%%%%%%%
\begin{figure}[t!]
    \centering
    \includegraphics[width=0.7\textwidth]{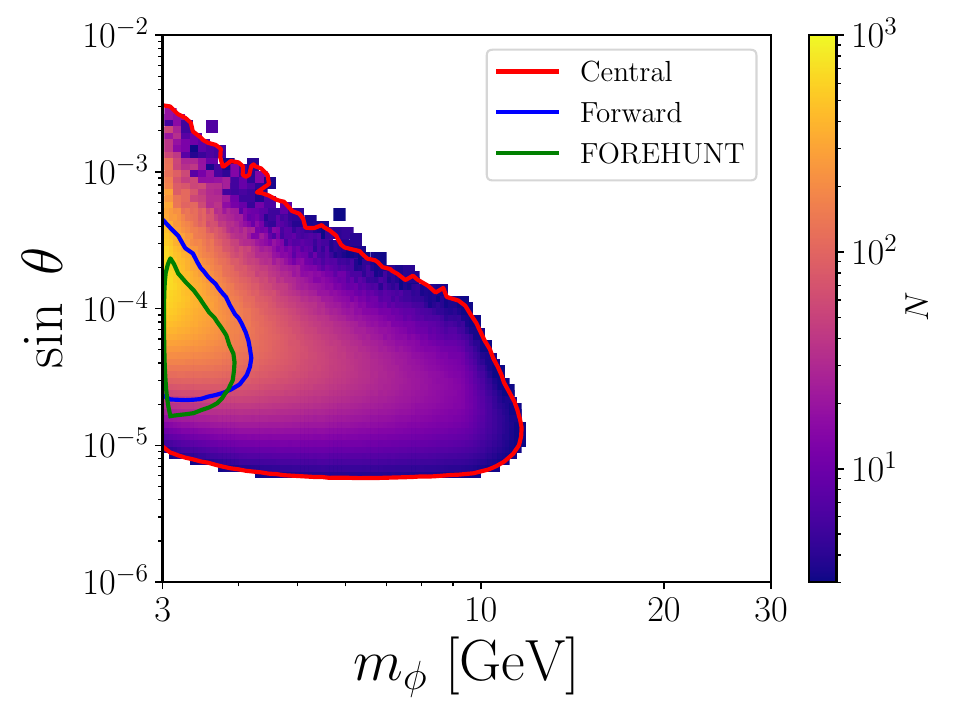}
    \caption{Number of signal events $N$ expected in the FCC Central tracker, in the plane $[m_\phi,\, \sin\theta]$, for an integrated luminosity $\mathcal{L} = 20$~ab$^{-1}$. The red, blue and green lines correspond to $N = 3$ for the Central tracker, the Forward tracker and FOREHUNT, respectively.    }
    \label{fig:collider_exc}
\end{figure}
%%%%%%%%%%%%%%%%%%%%%%%%%%%%%%%%%%%%%%%%%%%%%%%%%%%%%%%%%%
With these acceptances, we can now compute the expected number of signal events $N$ at a given detector as
\begin{equation} \label{eq:N_events}
    N = \sigma(p p \to \phi) \times \text{BR}(\phi \to f \bar{f}) \times \mathcal{L} \times \langle\epsilon\rangle\,.
\end{equation}

Figure~\ref{fig:collider_exc} shows the number of signal events expected in the FCC Central tracker, in the plane $[m_\phi,\, \sin\theta]$, for an integrated luminosity $\mathcal{L} = 20$~ab$^{-1}$. A realistic background estimate would require a detector-level study. Potential backgrounds include heavy-flavor decays, material interactions, long-lived SM hadrons, pile-up and reconstruction fakes. We do not model these backgrounds here and therefore interpret the $N = 3$ contour, used as a benchmark 95\% C.L. sensitivity criterion (red contour), as an idealized zero-background benchmark~\cite{Ellis:2012xd}. The figure also shows the regions corresponding to $N = 3$ for the FCC Forward tracker and FOREHUNT in blue and green, respectively. The reach of these two projects is very similar, with FOREHUNT extending its reach to lower values for the mixing $\sin\theta$ because it is further away in the forward region. As expected, the Central tracker offers the highest sensitivity as a result of its larger acceptance.

%%%%%%%%%%%%%%%%%%%%%%%%%%%%%%%%%%%%%
\section{Cosmology at colliders} \label{sec:results}
%%%%%%%%%%%%%%%%%%%%%%%%%%%%%%%%%%%%%
We now translate the FCC-hh displaced-decay reach into a probe of the early cosmological history. Throughout this work, we have assumed that the scalar $\phi$ dominated the energy density of the Universe before decaying into SM radiation. The temperature $\Trh$ therefore denotes the temperature of the SM plasma at the end of this $\phi$-dominated era. The connection with collider searches is direct. The Higgs mixing angle controls both the production rate of $\phi$ at colliders and its decay rate in the early Universe. Small values of $\sin\theta$ make $\phi$ long-lived, leading to displaced decays at FCC-hh, while also delaying the transition to radiation domination. Therefore, a displaced-decay search in the $[m_\phi,\, \sin\theta]$ plane can be reinterpreted as a probe of the $[m_\phi,\, \Trh]$ plane by using Eq.\eqref{eq:sintheta_Trh_thermal}.

%%%%%%%%%%%%%%%%%%%%%%%%%%%%%%%%%%%%%%%%%%%%%%%%%%%%%%%%%%
\begin{figure}[t!]
    \centering
    \includegraphics[width=0.495\textwidth]{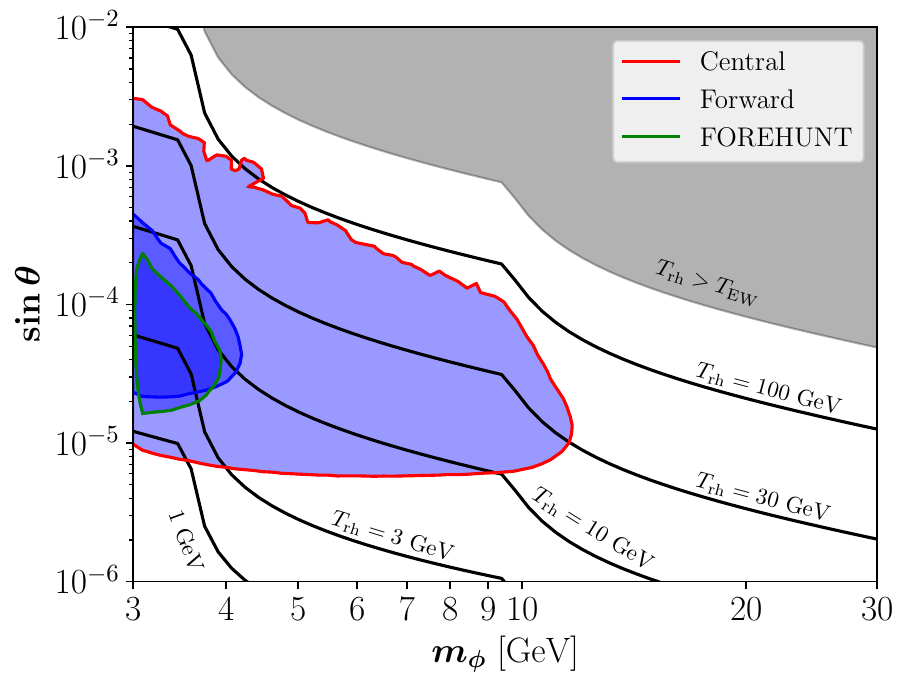}
    \includegraphics[width=0.495\textwidth]{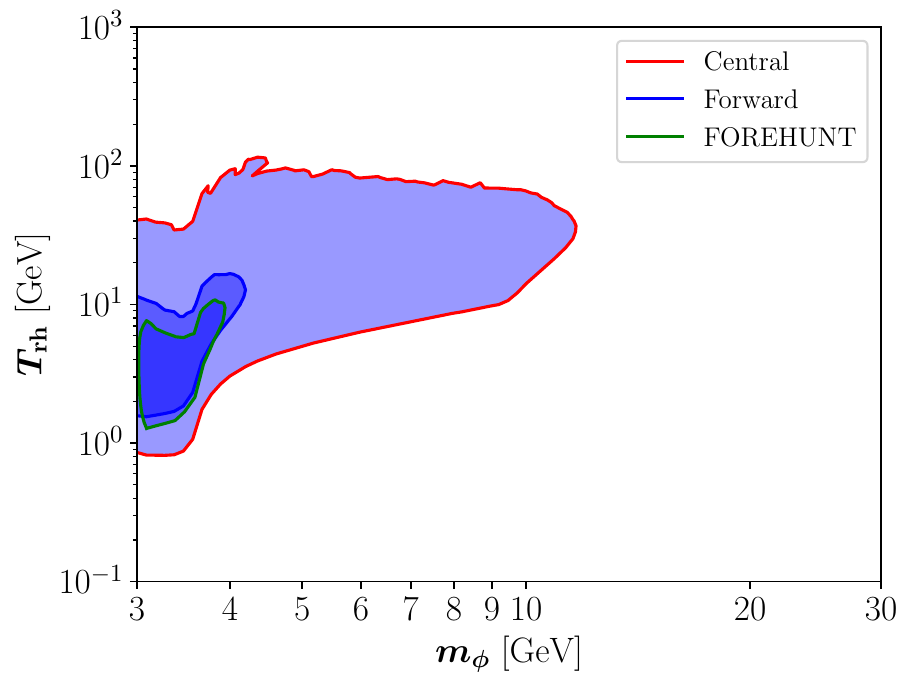}
    \caption{FCC-hh sensitivity to $\phi$ for an integrated luminosity $\mathcal{L} = 20$~ab$^{-1}$, overlaid with the reheating temperature $\Trh$. Left: projected displaced-decay reach in the $[m_\phi,\, \sin\theta]$ plane for the Central tracker, Forward tracker and FOREHUNT, together with contours of constant $\Trh$. Right: the same reach mapped onto the $[m_\phi,\, \Trh]$ plane. The part of the mapped reach with $\Trh > T_{\rm EW}$ is shown only as a formal extrapolation; the present interpretation is restricted to $\Trh < T_{\rm EW}$.}
    \label{fig:all}
\end{figure}
%%%%%%%%%%%%%%%%%%%%%%%%%%%%%%%%%%%%%%%%%%%%%%%%%%%%%%%%%%
This mapping is shown in Fig.~\ref{fig:all}. In the left panel, the projected FCC-hh sensitivity regions from Fig.~\ref{fig:collider_exc} are overlaid with contours of constant $\Trh$. The central tracker gives the largest coverage, due to its larger geometric acceptance and its proximity to the interaction point. The forward tracker and FOREHUNT have smaller acceptances in the region where the decay length is short, but they extend the sensitivity toward smaller mixing angles and longer lifetimes. The reach is concentrated around 3~GeV $\lesssim m_\phi \lesssim 15$~GeV and $6 \times 10^{-6} \lesssim  \sin\theta \lesssim 3 \times 10^{-3}$, with the exact boundary depending on the detector geometry.

The right panel of Fig.~\ref{fig:all} shows the same reach directly in the $[m_\phi,\, \Trh]$ plane. Within the region where the broken-phase Higgs-portal description is valid, the FCC-hh could probe transition temperatures from around the GeV scale up to the electroweak scale. The increase in $\Trh$ around $m_\phi \simeq 3.5$~GeV reflects the kinematical threshold of the $\tau$ leptons (cf. Fig.~\ref{fig:dec}). This is the most robust cosmological interpretation of our result. In terms of reheating temperature, we note that the FCC-hh reach lies well above the MeV scale, implying that the collider-accessible region corresponds to early matter-dominated eras that end safely before BBN, but below the electroweak scale.

Figure~\ref{fig:all} therefore illustrates the main message of this work: displaced decays of a light Higgs-portal scalar at FCC-hh can be interpreted as a probe of the pre-BBN expansion history. In the minimal reheating interpretation, the FCC-hh would test low-temperature Higgs-portal reheating. In a more general interpretation, it would probe the decay of a long-lived matter component that ended an early matter-dominated era and initiated radiation domination.

%%%%%%%%%%%%%%%%%%%%%%%%%%%%%%%%%%%%%%%%%%%%%%%%%%%%%%%%%%
\section{Conclusions} \label{sec:concl}
%%%%%%%%%%%%%%%%%%%%%%%%%%%%%%%%%%%%%%%%%%%%%%%%%%%%%%%%%%
The pre-BBN thermal history of the Universe may have included an early matter-dominated era driven by a long-lived particle or coherent scalar field. In this work, we studied a minimal Higgs-portal realization of this possibility, where a real singlet scalar $\phi$ mixes with the SM Higgs boson and later decays into SM states. We have remained agnostic about the origin of the primordial abundance of $\phi$. The simplest interpretation is ordinary cosmic reheating, with $\phi$ identified as the inflaton or reheaton, but the same framework also applies to more general matter-dominated epochs, such as those driven by moduli-like fields.

In this setup, the Higgs mixing angle controls both the early-Universe decay rate and the collider production rate of $\phi$. For small mixing, the scalar can be long-lived on collider scales while also decaying late enough to set the transition temperature $\Trh$, defined here as the temperature at which $\phi$-domination ends and SM radiation domination begins. We included the finite-temperature suppression of the decay width,
\[
    \Gp(T) = \Gp^{(0)}\, \tanh\!\left(\frac{m_\phi}{4\, T}\right),
\]
which arises from Pauli blocking and inverse decays when \(\phi\) decays into thermalized SM fermions. This effect lowers the transition temperature relative to the usual vacuum-decay estimate and is therefore important when translating collider-accessible lifetimes into early-Universe histories.

We found that the LHC has limited sensitivity to the GeV-scale region of interest, mainly because the small mixing suppresses production and because existing displaced-vertex searches are not optimized for light LLPs with soft visible decay products. The FCC-hh, with its larger center-of-mass energy and integrated luminosity, offers a more promising probe. Using displaced-decay acceptances for the FCC-hh central tracker, a forward tracker, and FOREHUNT, we showed that future LLP searches could test Higgs-portal scalars in the approximate range
\[
    3~{\rm GeV}\lesssim m_\phi\lesssim 15~{\rm GeV}, \qquad 6 \times 10^{-6} \lesssim  \sin\theta \lesssim 3 \times 10^{-3}.
\]
Within the broken-phase Higgs-portal description, this corresponds to transition temperatures from around the GeV scale up to the electroweak scale.

Our results illustrate that displaced-decay searches at future colliders can probe not only weakly coupled new particles, but also the expansion history of the Universe before BBN. In the minimal interpretation, the FCC-hh would test low-temperature Higgs-portal reheating. More generally, it would probe the end of an early matter-dominated era. A more complete treatment of the cosmological Boltzmann evolution, finite-temperature effects near thresholds, and realistic detector efficiencies and backgrounds would further sharpen this connection.

%%%%%%%%%%%%%%%%%%%%%%%%%%%%%%%%%%%%%%%%%%%
\acknowledgments
%%%%%%%%%%%%%%%%%%%%%%%%%%%%%%%%%%%%%%%%%%%
NB received funding from the grants PID2023-151418NB-I00 funded by MCIU/AEI/10.13039 /501100011033/ FEDER and PID2022-139841NB-I00 of MICIU/AEI/10.13039/501100011033 and FEDER, UE. GC acknowledges support from ANID FONDECYT grant No. 1250135 and ANID – Millennium Science Initiative Program ICN2019\_044. 

%%%%%%%%%%%%%%%%%%%%%%%%%%%%%%%%%%%%%%%
\appendix
%%%%%%%%%%%%%%%%%%%%%%%%%%%%%%%%%%%%%%%%%%
%%%%%%%%%%%%%%%%%%%%%%%%%%%%%%%%%%%%%%%%%%%%%%%%%%%%%%%%%%%%%%%%%%%%%%%%%%%%%%%%%%%%%%
\section{Decay in a thermal plasma} \label{sec:decay} 
%%%%%%%%%%%%%%%%%%%%%%%%%%%%%%%%%%%%%%%%%%%%%%%%%%%%%%%%%%%%%%%%%%%%%%%%%%%%%%%%%%%%%%
The particle $\phi$ with mass $m_\phi$ decays into a pair of fermions $\psi\, \bar\psi$, with a constant squared matrix element $|{\cal M}|^2$. The {\it in-vacuum} decay width $\Gp^{(0)}$ of $\phi$ is given by
\begin{equation}
    \Gp^{(0)} = \frac{|{\cal M}|^2}{16\pi\, m_\phi}\,,
\end{equation}
ignoring the masses of the final-state particles. However, in the early Universe $\phi$ does not decay in vacuum, but in the presence of a SM thermal plasma, which can have an impact on its cosmic evolution. Therefore, to understand the decay of $\phi$, we have to follow its phase-space distribution.

The evolution of the phase-space distribution function $f_\phi(t, p_\phi)$ of $\phi$, with momentum $p_\phi$, is given by the {\it unintegrated} Boltzmann equation
\begin{equation} \label{eq:unintegrated}
    \left(\frac{\partial}{\partial t} - H\, p_\phi\, \frac{\partial}{\partial p_\phi}\right) f_\phi(t, p_\phi) = \mathcal{C}[f_\phi]\,,
\end{equation}
where $\mathcal{C}[f_\phi]$ is the collision term defined as
\begin{align} \label{eq:collision}
    {\cal C}[f_\phi ] &\equiv \frac{|{\cal M}|^2}{2 E_\phi} \int d\Pi_2\, d\Pi_3\, (2\pi)^4\, \delta^{(4)}(p_\phi - p_2 - p_3) \left[-f_\phi \left(1 - f_2\right) \left(1 -  f_3\right) + f_2\, f_3 \left(1 + f_\phi \right)\right] \nonumber\\
    &= \frac{|{\cal M}|^2}{2 E_\phi} \int d\Pi_2\, d\Pi_3\, (2\pi)^4\, \delta^{(4)}(p_\phi - p_2 - p_3) \left[-f_\phi + 2 f_\phi\, f_2 + f_2\, f_3\right]  \nonumber\\
    &= \mathcal{C}_A + \mathcal{C}_B + \mathcal{C}_C\,,
\end{align}
the subscripts $2$ and $3$ denote quantities for the two fermions in the final state, $E_\phi = \sqrt{m_\phi^2 + p_\phi^2}$ is the energy of $\phi$, and $d\Pi_i$ is the Lorentz-invariant phase space for the particle $i$. In addition, we assume fermions $\psi$ to be thermally in equilibrium with the SM bath at a temperature $T$, so that their phase-space distribution is given by the Fermi-Dirac distribution $f_2(p) = f_3(p) = \frac{1}{\exp(p/T) + 1}$. In Eq.~\eqref{eq:collision}, the term proportional to $-f_\phi \left(1 - f_2\right) \left(1 -  f_3\right)$ corresponds to the decay with the corresponding Fermi-suppression factors, while the term proportional to $f_2\, f_3 \left(1 + f_\phi \right)$ comes from the inverse decay with its Bose-enhancement. We encounter the following phase-space integrals that can be solved following e.g. Ref.~\cite{Bernal:2026dsu}
\begin{align}
    \mathcal{C}_A &\equiv - \frac{|{\cal M}|^2}{2E_\phi}\int d\Pi_2\,d\Pi_3\,(2\pi)^4\,\delta^{(4)}(p_\phi -p_2-p_3)\,f_\phi \nonumber \\
    &= -\frac{|{\cal M}|^2}{2E_\phi} f_\phi \int \frac{d^3\vec p_2}{2 E_2 (2\pi)^3}\, \frac{d^3\vec p_3}{2 E_3 (2\pi)^3}\,(2\pi)^4\, \delta(E_\phi -E_2-E_3)\, \delta^{(3)}(\vec p_\phi - \vec p_2 - \vec p_3) \nonumber \\
    &= -\frac{|{\cal M}|^2}{32\pi^2E_\phi}f_\phi \left.\int\frac{d^{3}\vec{p}_2}{E_2}\,\frac{1}{E_3}\,\delta(E_\phi-E_2-E_3)\right|_{E_3=\sqrt{p_\phi^2+p_2^2-2p_\phi p_2c_\theta}}\nonumber \\
    &= -\frac{|{\cal M}|^2}{32\pi^2E_\phi}f_\phi \int\frac{2\pi\,p_2^2\,dp_2\,dc_\theta}{p_2}\,\frac{\delta\left(E_\phi-E_2-\sqrt{p_\phi^2+p_2^2-2p_\phi p_2c_\theta}\right)}{\sqrt{p_\phi^2+p_2^2-2p_\phi p_2c_\theta}}\nonumber \\
    &= -\frac{|{\cal M}|^2}{16\pi\,E_\phi}f_\phi \int\frac{p_2^2\,dp_2}{p_2}\,\frac{1}{p_\phi \,p_2}\,\Theta\left[\frac{E_\phi-p_\phi }{2}\leq p_2\leq\frac{E_\phi+p_\phi }{2}\right] = -\frac{|{\cal M}|^2}{16 \pi\, E_\phi}\, f_\phi \,,\label{eq:-66}
\end{align}
where $c_\theta$ is the cosine of the angle between the relevant momenta in the phase-space integral. Next, we compute $\mathcal{C}_B$, which is similar to $\mathcal{C}_A $, except for an additional factor of $f_2$
\begin{align}
    \mathcal{C}_B &\equiv 2\times \frac{|{\cal M}|^2}{2E_\phi} f_\phi \int d\Pi_2\,d\Pi_3\,(2\pi)^4\,\delta^{(4)}(p_\phi -p_2-p_3)\,f_2(p_2)\nonumber \\
    & =2\,\frac{|{\cal M}|^2}{16\pi\,E_\phi}f_\phi \int\frac{p_2^2\,dp_2}{p_2}\,\frac{f_2(p_2)}{p_\phi \,p_2}\, \Theta\left[\frac{E_\phi-p_\phi }{2}\leq p_2\leq\frac{E_\phi+p_\phi }{2}\right]\nonumber \\
    & =\frac{2\, |{\cal M}|^2}{16 \pi\, E_\phi}\, f_\phi\, \frac{1}{p_\phi}\int_{\frac{E_\phi-p_\phi }{2}}^{\frac{E_\phi+p_\phi }{2}}dp_2\, \frac{1}{e^{p_2/T} + 1} = \frac{|{\cal M}|^2}{16\pi\,E_\phi}\, f_\phi \left[2 + \frac{2\, T}{p_\phi}\, \ln\left(\frac{1 + e^\frac{E_\phi - p_\phi}{2 T}}{1 + e^\frac{E_\phi + p_\phi}{2 T}}\right)\right].
\end{align}
Lastly, $\mathcal{C}_C$ leads to
\begin{align}
    \mathcal{C}_C &= \frac{|{\cal M}|^2}{2E_\phi} \int d\Pi_2\,d\Pi_3\,(2\pi)^4\,\delta^{(4)}(p_\phi -p_2-p_3)f_2f_3\,,\nonumber \\
    &= \frac{|{\cal M}|^2}{32\pi^2\,E_\phi}\int\frac{d^{3}\vec{p}_2}{E_2}\frac{d^{3}\vec{p}_3}{E_3}\,f_2 (p_2)\,f_3(p_3)\,\delta(E_\phi-E_2-E_3)\,\delta^{(3)}(\vec{p}_\phi -\vec{p}_2-\vec{p}_3)\nonumber \\
    &= \frac{|{\cal M}|^2}{32\pi^2\,E_\phi}\left.\int\frac{d^{3}\vec{p}_2}{E_2}\,\frac{f_2 (p_2)\,f_3(p_3)}{E_3}\,\delta(E_\phi-E_2-E_3)\right|_{\vec{p}_3=\vec{p}_\phi -\vec{p}_2}\nonumber \\
    &= \frac{|{\cal M}|^2}{32\pi^2\, E_\phi} \int\frac{2 \pi\, p_2^2\, dp_2\, dc_\theta}{p_2}\, \frac{f_2 (p_2)\,f_3(E_\phi - p_2)}{p_2\, p_\phi}\, \delta\left(c_\theta - \frac{p_\phi^2 - E_\phi^2 + 2 E_\phi\, p_2}{2\, p_\phi\, p_2}\right) \nonumber\\
    &= \frac{|{\cal M}|^2}{16\pi\, E_\phi}\, \frac{1}{p_\phi} \int dp_2\, \frac{1}{e^{p_2/T} + 1}\, \frac{1}{e^{(E_\phi - p_2)/T} + 1}\, \Theta\left[\frac{E_\phi-p_\phi }{2}\leq p_2\leq\frac{E_\phi+p_\phi }{2}\right]\nonumber\\
    &= \frac{|{\cal M}|^2}{16\pi\, E_\phi}\, \frac{2\, T}{p_\phi}\, \frac{1}{e^{E_\phi/T} - 1}\, \ln\left[\frac{e^\frac{E_\phi + p_\phi}{4 T} + e^{-\frac{E_\phi + p_\phi}{4 T}}}{e^\frac{E_\phi - p_\phi}{4 T} + e^{-\frac{E_\phi - p_\phi}{4 T}}}\right] \nonumber\\
    &= \frac{|{\cal M}|^2}{16\pi\, E_\phi}\, f_\phi^\text{eq} \left[-1 + \frac{2\, T}{p_\phi}\, \ln\left(\frac{1 + e^\frac{E_\phi + p_\phi}{2 T}}{1 + e^\frac{E_\phi - p_\phi}{2 T}}\right)\right],
\end{align}
where $f_\phi^\text{eq}(p_\phi) \equiv \frac{1}{\exp(E_\phi/T) - 1}$ is the Bose-Einstein equilibrium distribution for $\phi$. The collision term in Eq.~\eqref{eq:collision} then becomes
\begin{equation}
    \mathcal{C}[f_\phi] = \frac{|{\cal M}|^2}{16\pi\,E_\phi} \left[1 + \frac{2\, T}{p_\phi}\, \ln\left(\frac{1 + e^\frac{E_\phi - p_\phi}{2 T}}{1 + e^\frac{E_\phi + p_\phi}{2 T}}\right)\right] \left(f_\phi(p_\phi) -f_\phi^\text{eq}(p_\phi) \right).
\end{equation}
We emphasize that this result assumes that the fermions in the final state are in full thermal equilibrium (that is, they are described by a thermal distribution with zero chemical potential) and negligible thermal masses.

The {\it integrated} collision terms can be analytically solved in the {\it non-relativistic} limit for the inflaton, and are
\begin{align}
    \int \frac{d^3\vec p_\phi}{(2\pi)^3}\, \mathcal{C}_A &\simeq - \frac{|{\cal M}|^2}{16\pi\, m_\phi}\, \int \frac{d^3\vec p_\phi}{(2\pi)^3}\, f_\phi (p_\phi) = -\Gp^{(0)}\, n_\phi\,,\\
    \int \frac{d^3\vec p_\phi}{(2\pi)^3}\, \mathcal{C}_B &\simeq \frac{|{\cal M}|^2}{8\pi\, m_\phi}\, \int \frac{d^3\vec p_\phi}{(2\pi)^3}\, f_\phi(p_\phi)\, \frac{1}{e^\frac{m_\phi}{2 T} + 1} = 2\, \Gp^{(0)}\, n_\phi\, \mathcal{F}\,,\\
    \int \frac{d^3\vec p_\phi}{(2\pi)^3}\, \mathcal{C}_C &\simeq n_\phi^\text{eq}\, \Gp^{(0)} \left(1 - 2\, \mathcal{F}\right),
\end{align}
where
\begin{equation}
    \mathcal{F} \equiv \frac{1}{e^\frac{m_\phi}{2 T} + 1}\,.
\end{equation}

The first moment of the Boltzmann equation~\eqref{eq:unintegrated} corresponds to the evolution of the number density $n_\phi$ of $\phi$, and is given by~\cite{Adshead:2019uwj}
\begin{equation}
    \frac{dn_\phi}{dt} + 3\, H\, n_\phi = \int \frac{d^3\vec p_\phi}{(2\pi)^3}\, \mathcal{C}[f_\phi] \simeq - \Gp(T) \left(n_\phi - n_\phi^\text{eq}\right),
\end{equation}
where
\begin{equation} \label{eq:Geff}
    \Gp(T) \equiv \Gp^{(0)} \left(1 - 2\, \mathcal{F}\right) = \Gp^{(0)}\, \tanh\left(\frac{m_\phi}{4\, T}\right),
\end{equation}
where $\Gp(T)$ corresponds to the {\it in-medium} temperature-dependent decay width of $\phi$. We emphasize that the hyperbolic tangent in Eq.~\eqref{eq:Geff} corresponds to the backreaction of the SM plasma to the decay of $\phi$, due to quantum statistics in the decay. Hence, $\Gp(T)$ can be suppressed with respect to the case of decay in vacuum $\Gp^{(0)} = \Gp(T = 0)$, increasing the lifetime of $\phi$. We note that Eq.~\eqref{eq:Geff} can be approximated as 
\begin{equation}
    \Gp(T) \simeq \Gp^{(0)} \times
    \begin{dcases}
        1 &\text{ for }T \leq \frac{m_\phi}{4}\,,\\
        \frac{m_\phi}{4\, T} &\text{ for }T \geq \frac{m_\phi}{4}\,.
    \end{dcases}
\end{equation}

Modification of the decay of $\phi$ has a strong cosmological impact. The reheating temperature $\Trh$ corresponds to the SM temperature at the moment of the equality of the energy densities of the SM and $\phi$. It can be estimated analytically by $H(\Trh) = \Gp(\Trh)$, where
\begin{equation}
    H(T) = \frac{\pi}{3} \sqrt{\frac{\gs(T)}{10}}\, \frac{T^2}{M_P}\,,
\end{equation}
for a universe dominated by SM radiation and where $\gs$ corresponds to the number of relativistic degrees of freedom that contribute to the SM energy density. Therefore,
\begin{equation}
    \Trh \simeq
    \begin{dcases}
        \left(\frac{3}{\pi}\right)^{1/2} \left(\frac{10}{\gs}\right)^{1/4} \left(M_P\, \Gp^{(0)}\right)^{1/2} &\text{ for } m_\phi \geq 4\, \Trh\,,\\
        \left(\frac{3}{4\pi}\right)^{1/3} \left(\frac{10}{\gs}\right)^{1/6} \left(M_P\, \Gp^{(0)}\, m_\phi\right)^{1/3} &\text{ for } m_\phi \leq 4\, \Trh\,.
    \end{dcases}
\end{equation}

%%%%%%%%%%%%%%%%%%%%%%%%%%%%%%%%%%%%%%%%%%%%%%
\section{Inflaton decay kinematics} \label{sec:appendix}
%%%%%%%%%%%%%%%%%%%%%%%%%%%%%%%%%%%%%%%%%%%%%%
%%%%%%%%%%%%%%%%%%%%%%%%%%%%%%%%%%%%%%%%%%%
\begin{figure}[t!]
    \centering
    \includegraphics[width=0.85\textwidth]{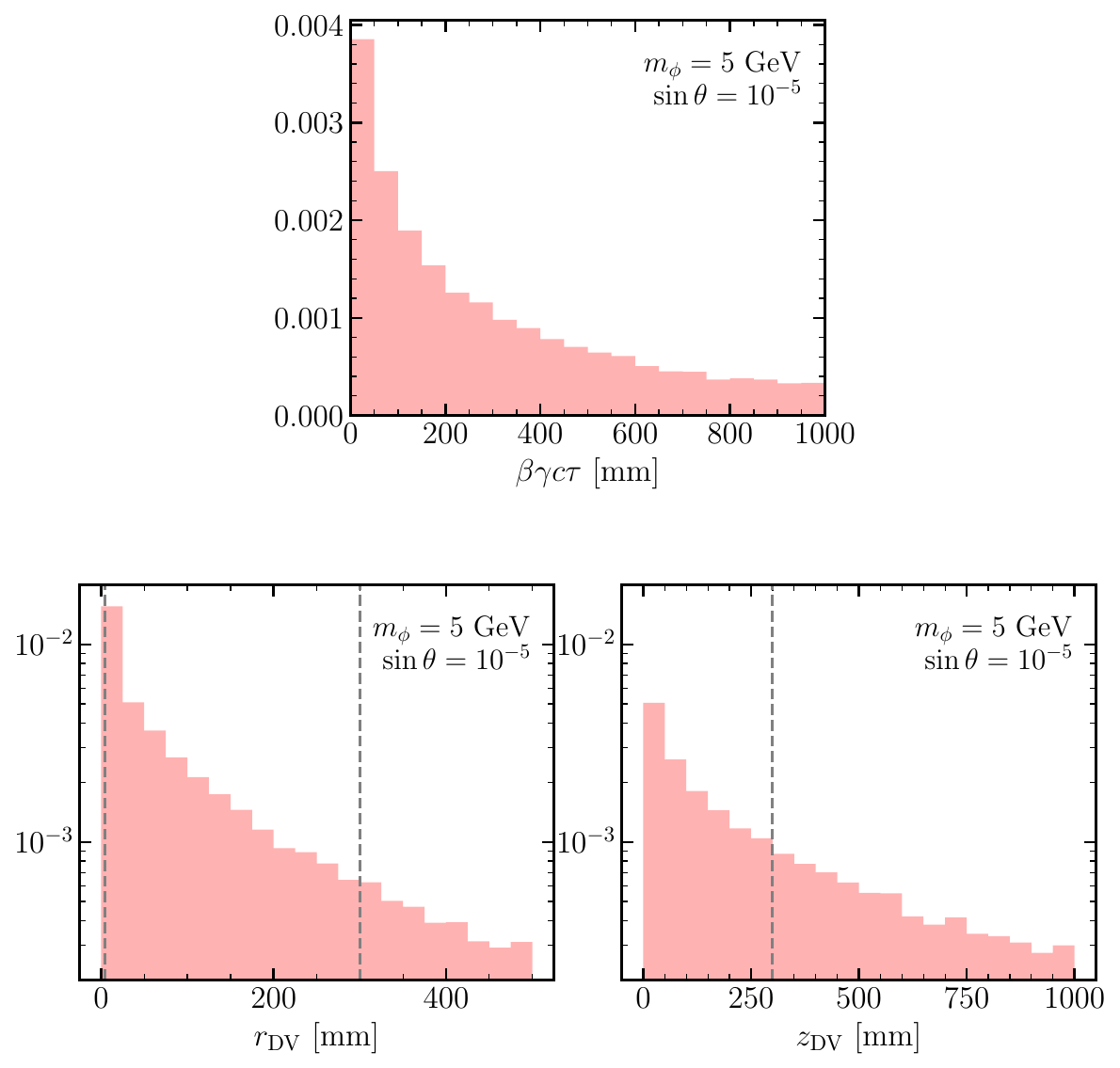}
    \caption{Boosted decay length for a fixed benchmark of ($m_\phi=5$~GeV, $\sin\theta=10^{-5}$) (top). Transverse (bottom left) and longitudinal (bottom right) decay distances of the DV. The vertical gray lines show the DV cuts of ATLAS search at 4 and 300 mm. }
    \label{fig:decay_dist}
\end{figure}
\begin{figure}[t!]
    \centering
    \includegraphics[width=0.85\textwidth]{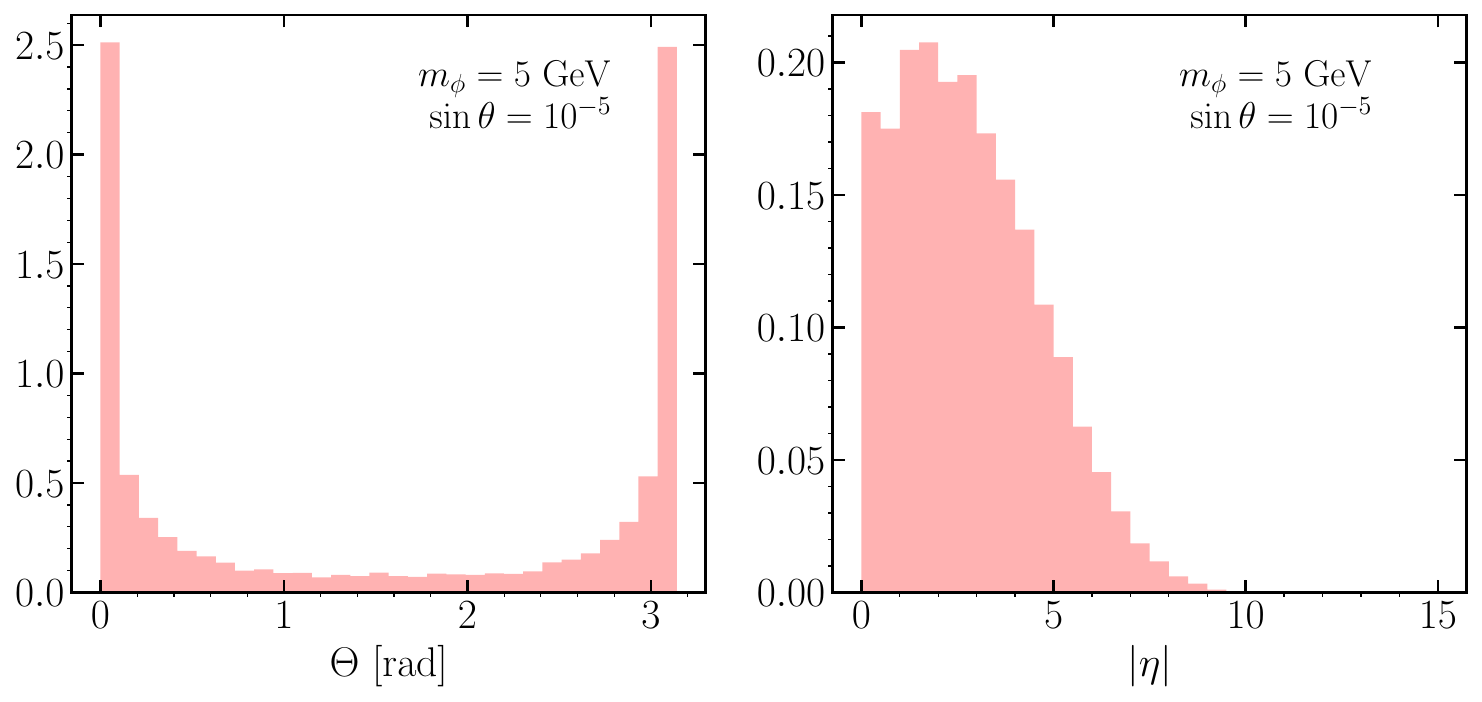}
    \caption{Polar angle $\Theta$ and pseudo-rapidity $|\eta|$ distributions for a fixed benchmark of ($m_\phi=5$~GeV, $\sin\theta=10^{-5}$).}
    \label{fig:eta5}
\end{figure}

%%%%%%%%%%%%%%%%%%%%%%%%%%%%%%%%%%%%%%%%%%%
We use $\texttt{Pythia 8}$~\cite{Sjostrand:2014zea} to extract kinematical distributions related to the displaced phenomenology of the scalar $\phi$ for two benchmarks: one falling in the bulk of the FCC sensitivity region with displaced decays ($m_\phi=5$~GeV and $\sin\theta=10^{-5}$) that escapes the ATLAS DV + jets search~\cite{ATLAS:2023oti} due to the invariant mass cut of the DV of 10~GeV; and one higher mass benchmark ($m_\phi=50$~GeV and $\sin\theta=10^{-5}$), but whose transverse ($r_{\rm DV}$) and longitudinal ($z_{\rm DV}$) decay positions of the displaced vertex falls outside the search cuts, as these require 4~mm $ < r_{\rm DV} < 300$~mm and $|z_{\rm DV}|< 300$~mm. Figure~\ref{fig:decay_dist} shows the boosted decay distance (top plot) and the transverse (bottom left) and longitudinal (bottom right) decay distances. Figure~\ref{fig:eta5} shows the polar angle $\Theta$ and the absolute value of pseudo-rapidity $|\eta|$ for this representative benchmark. We note that our inflaton tends to be in the forward region along the beam pipe (at $\Theta=0,\pi$), with a large number of events within the angular acceptance of the forward FCC tracker ($|\eta|< 3$).
%%%%%%%%%%%%%%%%%%%%%%%%%%%%%%%%%%%%%%%%%%%
%%%%%%%%%%%%%%%%%%%%%%%%%%%%%%%%%%%%%%%%%%%

%%%%%%%%%%%%%%%%%%%%%%%%%%%%%%%%%%%%%%%%%%%
\begin{figure}[ht!]
    \centering
    \includegraphics[width=0.85\textwidth]{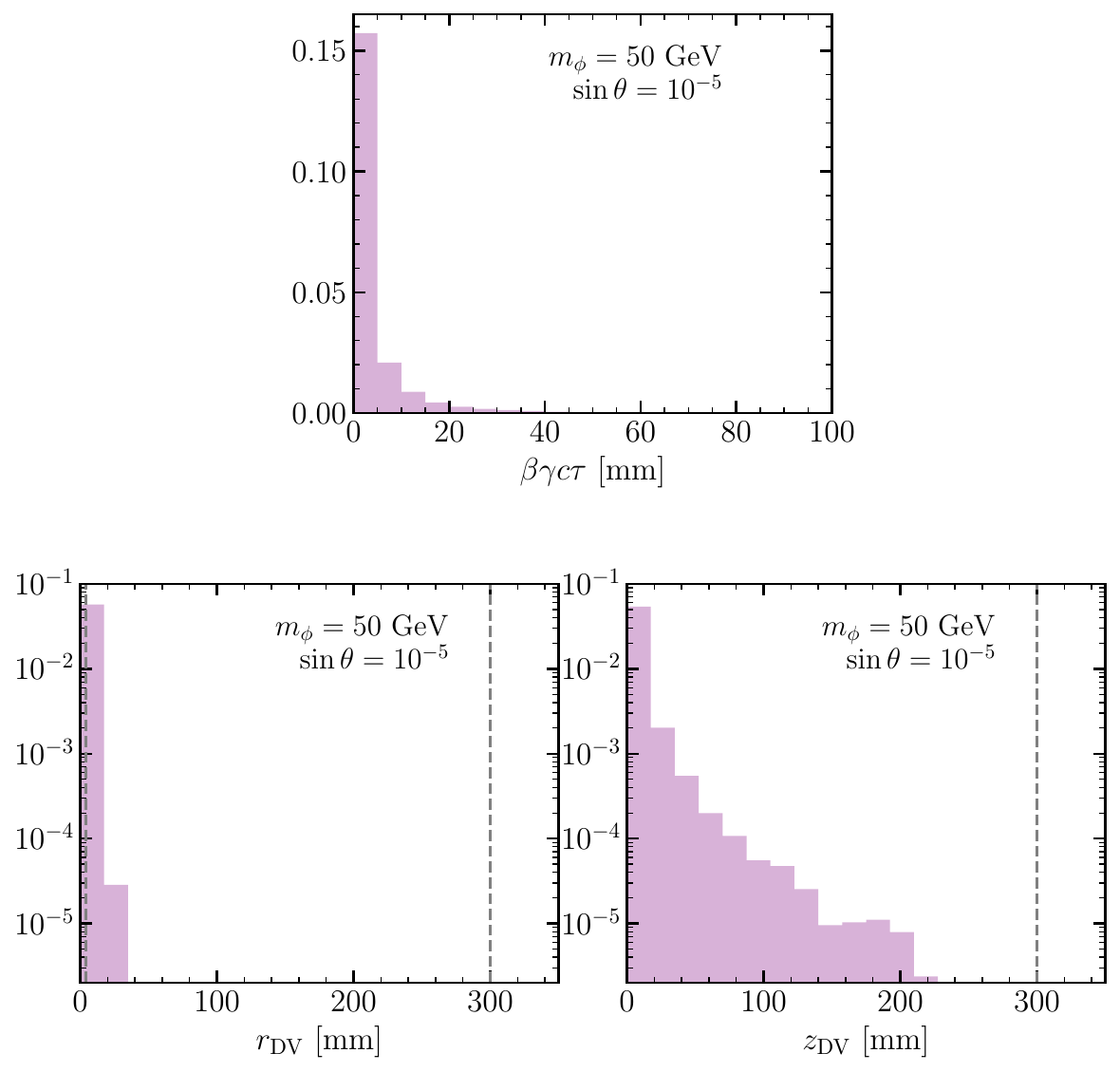}
     \caption{Boosted decay length for a fixed benchmark of ($m_\phi=50$~GeV, $\sin\theta=10^{-5}$) (top). Transverse (bottom left) and longitudinal (bottom right) decay distances of the DV. The vertical gray lines show the DV cuts of ATLAS search at 4 and 300mm.}
    \label{fig:decay_dist2}
\end{figure}
\begin{figure}[t!]
    \centering
        \includegraphics[width=0.85\textwidth]{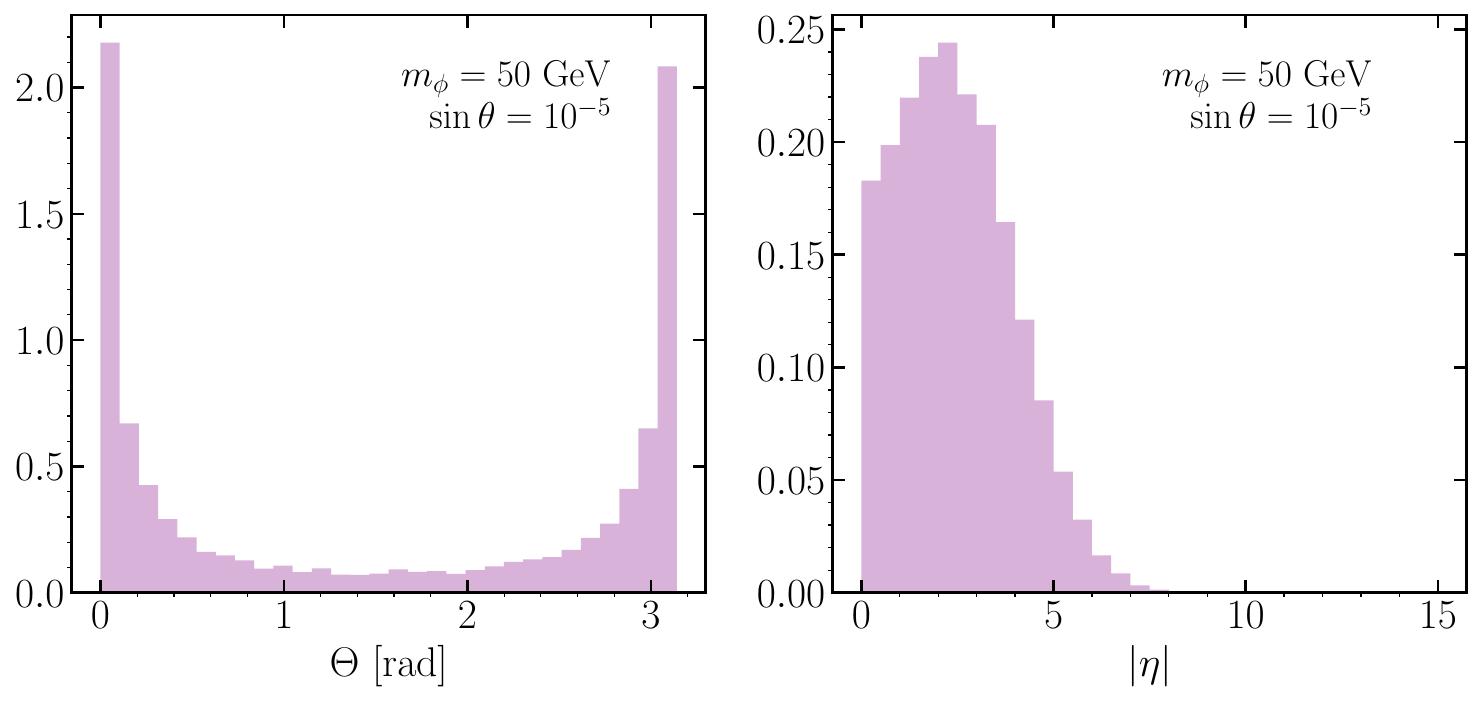}
    \caption{Polar angle $\Theta$ and pseudo-rapidity $|\eta|$ distributions for a fixed benchmark of ($m_\phi=50$~GeV, $\sin\theta=10^{-5}$).   }
    \label{fig:eta50}
\end{figure}

%%%%%%%%%%%%%%%%%%%%%%%%%%%%%%%%%%%%%%%%%%%
Figure~\ref{fig:decay_dist2} shows the boosted decay distance (top plot) and the transverse (bottom left) and longitudinal (bottom right) decay distances of a higher mass benchmark of $(m_\phi = 50$~GeV, $\sin\theta = 10^{-5})$, which would pass the DV mass cut but is already too prompt to pass the other cuts on transverse and longitudinal displacements of the ATLAS DV + jets search. Figure~\ref{fig:eta50} shows the polar angle $\Theta$ and the absolute value of pseudo-rapidity $|\eta|$ for this higher mass benchmark.
%%%%%%%%%%%%%%%%%%%%%%%%%%%%%%%%%%%%%%%%%%%
%%%%%%%%%%%%%%%%%%%%%%%%%%%%%%%%%%%%%%%%%%%

\begin{figure}[t!]
    \centering
    \includegraphics[width=0.85\textwidth]{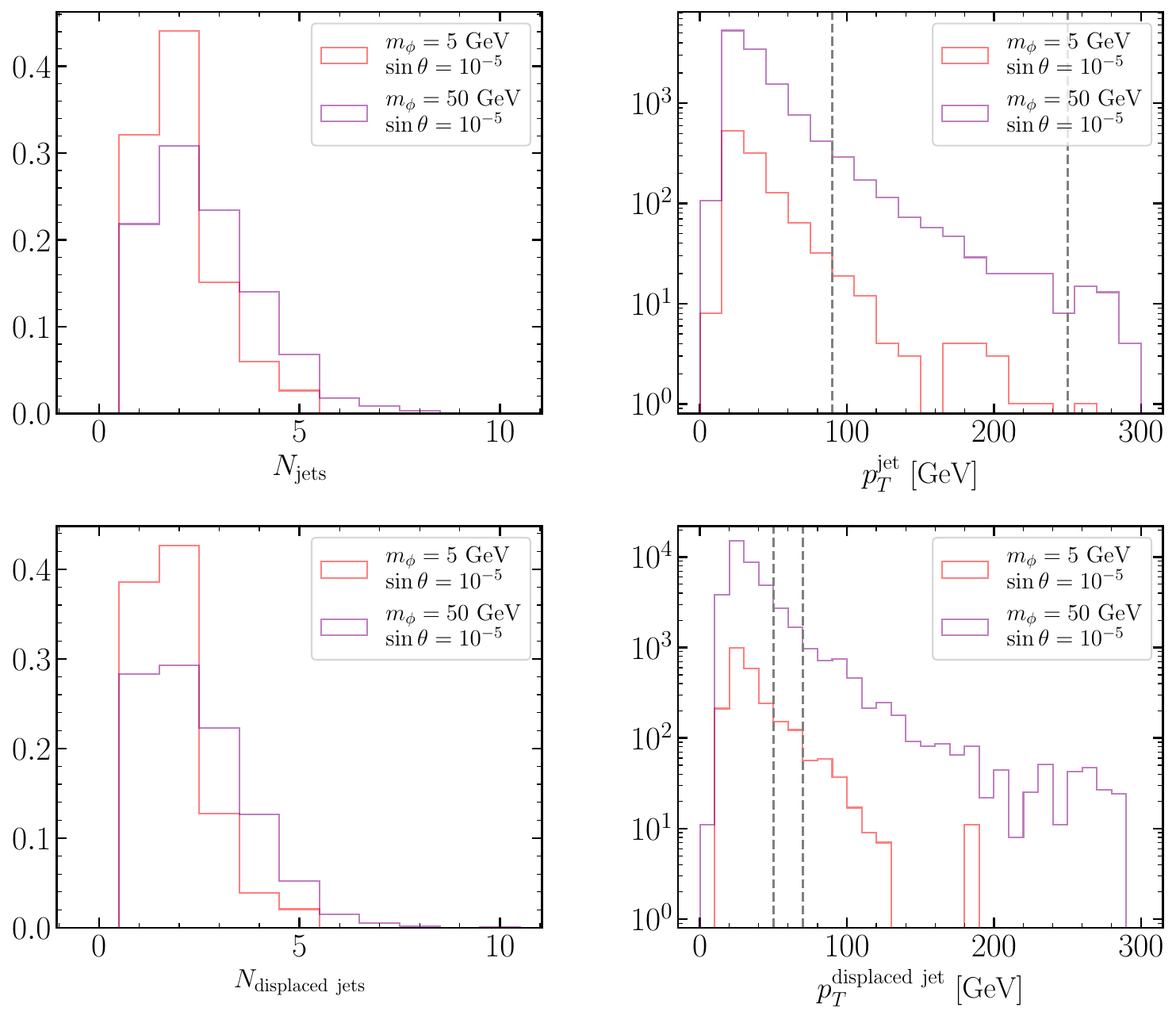}
    \caption{Number of jets and displaced jets (left column, normalized to unity) for two benchmarks with $(m_\phi=5$~GeV, $\sin\theta=10^{-5})$ (red) and $(m_\phi=50$~GeV, $\sin\theta=10^{-5})$ (purple), as well as their $p_T$ distributions (right column). Vertical lines show $p_T$ thresholds at 90~GeV and 250~GeV (top right plot) and of 50~GeV and 70~GeV (bottom left plot). }
    \label{fig:jets}
\end{figure}

We show for completeness the impact of the cuts placed on jets of the ATLAS DV + jet search in Fig.~\ref{fig:jets}. Following the recast performed by some of us in ref.~\cite{Cheung:2024qve}, we reconstruct truth-level jets and displaced jets (or ``trackless jets'' as defined in the ATLAS search~\cite{ATLAS:2023oti}) with \texttt{FastJet}~\cite{Cacciari:2011ma}, using the anti$-k_{t}$ algorithm with $R = 0.4$. Our jet definitions follow the instructions in the HEPData note for this analysis~\cite{ATLAS:2023oti}, where the truth jets are selected from all stable particles, including those from the LLP decay, excluding muons and neutrinos. Among these truth jets, displaced jets have the additional requirement that these match with a $\Delta R$ requirement of $\Delta R<0.3$ between the truth jet and the decay products of the LLP. Displaced truth jets also satisfy $|\eta| < 2.5$, and the matched LLPs should decay with a transverse distance from IP smaller than 3870 mm, corresponding to the region of the calorimeter.
%%%%%%%%%%%%%%%%%%%%%%%%%%%%%%%%%%%%%%%%%%%
%%%%%%%%%%%%%%%%%%%%%%%%%%%%%%%%%%%%%%%%%%%

Figure~\ref{fig:jets} shows the number of jets and displaced jets (normalized to unity) for two benchmarks, as well as their $p_T$ distributions. A peak is seen in 2 (displaced) jets, originating from the $\phi$ decay. The vertical lines in the right column of the figure represent the jet $p_T$ cuts of the search. For jets, these correspond to cuts on 90~GeV and 250~GeV, and high jet multiplicity is required. The loosest cuts for the jets correspond to at least 4 jets with $p_T>250$~GeV. or at least 7 jets with $p_T>90$~GeV. For displaced jets, the looser selections are at least 2 displaced jets with $p_T>50$~GeV or at least one displaced jet with $p_T>70$~GeV. This last selection is the most optimal for our most massive benchmark, but is already too prompt to pass the DV cuts (see Fig.~\ref{fig:decay_dist2}). 

In summary, we find that the searches for displaced vertex in ATLAS are not sensitive to our model parameter space due to strong cuts in the DV invariant mass, displaced decay positions, jet multiplicity and (displaced) jet $p_T$.

\clearpage

%%%%%%%%%%%%%%%%%%%%%%%%%
\bibliographystyle{JHEP}
\bibliography{biblio}
%%%%%%%%%%%%%%%%%%%%%%%%%
\end{document}